\documentclass[manuscript]{aastex}
\begin{document}
\title{Outer Architecture of Kepler-11: Constraints from Coplanarity}
\author{Daniel Jontof-Hutter\altaffilmark{1,2,3}, Brian P. Weaver\altaffilmark{2,3}, Eric B. Ford\altaffilmark{2,3}, Jack J. Lissauer\altaffilmark{4}, Daniel C. Fabrycky\altaffilmark{5}}
\email{djontofhutter@pacific.edu}
\altaffiltext{1}{Department of Physics, University of the Pacific, 3601 Pacific Avenue, Stockton, CA 95211, USA}
\altaffiltext{2}{Department of Astronomy, Pennsylvania State University, University Park, PA 16802, USA}
\altaffiltext{3}{Center for Exoplanets and Habitable Worlds, The Pennsylvania State University, University Park, PA 16802, USA}
\altaffiltext{4}{Space Science and Astrobiology Division, MS 245-3, NASA Ames Research Center, Moffett Field, CA 94035, USA}
\altaffiltext{5}{Department of Astronomy and Astrophysics, University of Chicago, 5640 South Ellis Avenue, Chicago, IL 60637, USA}

\begin{abstract}
The outer architectures of Kepler's compact systems of multiple transiting planets remain poorly constrained and few of these systems have lower bounds on the orbital distance of any massive outer planets. We infer a minimum orbital distance and upper limits on the inclination of a hypothetical Jovian-mass planet orbiting exterior to the six transiting planets at Kepler-11. Our constraints are derived from dynamical models together with observations provided by the \textit{Kepler} mission. Firstly, the lack of transit timing variations (TTV) in the outermost transiting planet Kepler-11 g imply that the system does not contain a Jovian mass perturber within 2 AU from the star. Secondly, we test under what initial conditions a Jovian-mass planet moderately inclined from the transiting planets would make their co-transiting configuration unlikely. The transiting planets are secularly coupled and exhibit small mutual inclinations over long time-scales, although the outermost transiting planet, Kepler-11 g is weakly coupled to the inner five. We rule out a Jovian-mass planet on a 3$^{\circ}$ inclination within 3.0 AU, and higher inclinations out to further orbital distances, unless an undetected planet exists orbiting in the dynamical gap between Kepler-11 f and Kepler-11 g. Our constraints depend little on whether we assume the six transiting planets of Kepler-11 were initially perfectly co-planar or whether a minimum initial mutual inclination between the transiting planets is adopted based on the measured impact parameters of the transiting planets.

\end{abstract}
\section{Introduction}
Kepler's discovery of multiple compact, coplanar systems of multi-transiting sub-Neptune planets has sparked renewed interest in planet formation theory. Kepler found 261 systems with three or more transiting planet candidates, and 92 with four or more transiting planets (retrieved from \textit{http://exoplanetarchive.ipac.caltech.edu}).  Most of these planets, super-Earths to sub-Neptunes are in a size range that is absent from the solar system, as are their remarkably compact configurations. Systems like Kepler-11, with six planets having orbital periods well under that of Venus \citep{liss11a}, highlight the difference between what appears common at other systems and our own Solar System. 

The debate over how these systems formed has, to a certain extent, given rise to two competing paradigms, invoking either the migration of multi-planet systems to close-in orbits \citep{Izidoro2015}, or in situ formation determined by local conditions (\citealt{chi13,Hansen2013,Petrovich2013}). Much of the discussion has occurred in the context of interpreting the period-ratio distribution of planet pairs, which is one of the major planet population observables delivered by the \textit{Kepler} mission.  

The period ratio distribution of Kepler's multi-transiting planets has narrow valleys and peaks near first-order mean motion resonances, over an otherwise smooth distribution (\citealt{liss11b, fab14, Steffen2015,Malhotra2015}). Generally, the assembly of resonances is thought to result from convergent migration and resonant trapping, followed by migration while maintaining constant orbital period ratios. This explains the preponderance of orbital period ratios near 2:1 or 3:2. However, closer orbital period ratios require either initial formation at a small orbital period ratio or a mechanism to avoid or escape from the 2:1 and 3:2 resonances. Furthermore, the migration of planets and resonant trapping leads to the expectation that the majority of multi-planet systems should end up in resonance or even in resonant chains, although that does not seem to be the case \citep{veras2012}. In fact, just narrow of first-order resonances, the period ratio distribution has narrow gaps, with sharp peaks just wide of the resonances (\citealt{liss11b,fab14}).

\citet{lithwu12} invoked tidal eccentricity damping of near resonant pairs to explain the gap and peak structure near resonances in the period ratio distribution. A similar result was found by \citet{Batygin2013}, who further suggested that many of the planets with near-commensurate orbital period ratios are in fact still in resonances. The tidal dissipation model requires rather efficient tidal damping (tidal Q$\sim$ 10), or a different mechanism beyond orbital periods of around 10 days, where tidal damping becomes less efficient. However, even with rocky planets among the sample, tides are not strong enough to move some of the planets to their observed separations \citep{Lee2013}. Nevertheless, the period ratio distribution changes significantly with orbital distance and the structure near resonance is more pronounced at shorter periods, which makes tidal dissipation a potentially important factor in explaining the near-resonant peaks \citep{Delisle2014}. 

An additional challenge to the migration model is that the majority of planet pairs are in fact not in or near resonance, but rather fill the period ratio distribution between the peaks and troughs at low order resonances. Various theories attempt to explain the period-ratio distribution include migration with a disk or planetesimal interactions. \citet{Rein2012} noted the difficulty of forming close first-order mean motion resonances with disk migration, and found that some fine tuning between smooth migration and stochastic interactions with turbulence or planetesimals can reproduce the observed period ratio distribution. Another explanation invokes the interaction between a planet and the wake of its neighbor within a disk, which can reverse convergent migration and increase the orbital period ratio away from resonance \citep{Baruteau2013}. Alternatively, \citet{Chatterjee2015} found that a disk of planetesimals can cause low mass planets to leave resonance after gas dispersal if the planetesimal disk mass is high enough. In their simulations, resonances persist among more massive planets.

\citet{Goldreich2014} invoked overstable librations in resonances to explain the low number of systems in resonance, where the eccentricities are high enough (given the dynamical masses of the planets) that resonant capture is only temporary.  

While there has been significant progress in reconciling the expectation of planets trapped in resonant chains with the observed period ratio distribution in migration models, \citet{Petrovich2013} showed that the characteristic peak-trough structure near resonances could also be a natural outcome of rapid in situ formation. Therefore, both in situ and migration models remain viable frameworks to explain the orbital period-ratios of \textit{Kepler}'s multi-transiting planets. Where the two paradigms differ most is in the outer architectures of the compact multi-transiting systems.

\citet{Izidoro2015} modelled the assembly of compact multi-planet systems as co-migrating ensembles of distantly formed Neptunes. In these models, the likelihood of the arrival of the Neptunes to distances within 1 AU depends on the presence of giant planets, which tend to block access to the inner regions. Under this scenario, the high-multiplicity systems of transiting planets should not have distant jovian planets.
 
On the other hand, in situ formation models allow planets to form in an extended disk over a wide range of distances in relative isolation. In such systems, the presence of a compact group of inner super-Earths or Neptunes would in no way preclude the formation of outer planets like the giant planets of our Solar System (\citealt{chi13,Hansen2013}).

Until an RV survey of compact multi-transiting planets is largely complete, we can only assess the potential effects of outer Jovian planets on the formation and architectures of these systems. \citet{BatyginLaughlin2015} considered the effect of the `Grand Tack' scenario \citep{Walsh2011}, whereby Jupiter migrates in as close as $\sim$1.5 AU before migrating outwards, trapped in resonance with Saturn. In a system like Kepler-11, such a scenario could drive a compact system of inner planets formed in situ into the star. Hence, a lack of outer Jovian planets or an alternative to a Grand Tack scenerio is predicted for such compact systems, if they formed in situ. 

For transiting systems with high multiplicity (four or more transiting planets), it will take some time before RV surveys can answer these questions. Very few high multiplicity systems have published RV constraints on the orbital architectures beyond the known transiting planets, due to the limited time available for observations, the large number of systems with four or more transiting planets and the faintness of many of the most interesting systems.  In the case of Kepler-11, RV monitoring with Keck HIRES since 2011 found the presence of Jovian-mass outer planet ($m\sin i = 1$ M$_{Jup}$) unlikely within an orbital period of 1000 days or $\approx 1.93$ AU \citep{Weiss2016}. 

However, dynamical models of the effects of distant Jovian planets on the observed transiting systems can reveal important insights into the outer architectures of multi-planet systems.  \citet{Hands2015} modeled the effects of outer Jovian planets on the assembly of resonances among compact multi-transiting systems like Kepler-11. They found that by disrupting wide, low-order resonances  (e.g., 2:1 and 3:2) between inner planets, massive Jovian companions enhance the likelihood of inner planets becoming trapped in closer resonances (e.g.,  4:3 and 5:4), or Laplace-like resonance chains. They predict that distant Jovians may be more likely at systems like Kepler-36, which contains transiting planets with a period ratio smaller than 6:5.

\citet{Hansen2016} considered the effects of secular perturbations on the multiplicity of transiting systems in the \textit{Kepler} dataset. They found that the excess of single transiting planets cannot be explained by the pumping of inclination without driving a significant fraction of systems to be dynamically unstable. A similar conclusion was drawn by \citet{Lai2016}. In their results, the maximum mutual inclination between two close-in planets in the presence of an external perturber is constrained by the mass ratio of the inner planets and their orbital separation. They found that, given a moderately inclined outer planet, the mutual inclination of a close-in pair increases with $m_p/a_p^3$, where $m_p$ and $a_p$ are the mass and orbital distance of the outer perturbing planet. 

In the case of Kepler-11, upper limits on the mass of a perturbing planet beyond the outermost transiting planet can be derived from two lines of inquiry with dynamical simulations, which may applicable to several systems of high multiplicity discovered by \textit{Kepler}. 

The first is to test the extent to which the existing transit timing data can rule out a Jovian companion beyond the transiting planets. In the case of Kepler-11, observed transit timing variations (TTVs) place useful upper limits on non-transiting perturbers that are near a low-order mean motion resonances with Kepler-11 g, or near enough to cause a detectable ``synodic chopping". We explain the method and results of this in the next section.  

The second line of inquiry is to test the effect of a distant Jupiter with a moderately inclined orbit on the remarkably co-planar configuration of Kepler-11 b-g. A similar analysis of the Kepler-20 and WASP-47 systems has been performed by \citet{Becker2017}. In the case of Kepler-11, the coupling between the six transiting planets can cause non-linear effects in mutual inclinations due to overlapping resonances. Nevertheless, \citet{Lai2016} have demonstrated that to a large extent the maximum mutual inclination of an inner system of two planets with an inclined perturber can be determined analytically, particularly in limit of strong secular coupling. Furthermore, the nonlinear effect in systems of higher multiplicity may be neglected in the strong coupling limit if the angular momentum of the inner system is dominated by one planet.  In the case of Kepler-11, five of the six known planets have tightly constrained masses, and their masses span much less than one order of magnitude \citep{liss13}. The inner five are all strongly coupled and their mutual inclinations remain small in the presence of a Jovian perturber. However, the wide seperation between Kepler-11 f and g leave this pair moderately coupled, and in this case, the amplitude of mutual inclination cycles is sensitive to the ratio of planetary masses among the inner pair. 

We refer to the transiting planets of Kepler-11 as the ``confirmed" planets throughout (even though Kepler-11 g is technically ``validated" by a probabilistic argument rather than being ``confirmed" by an alternative signature like TTVs, see \citealt{liss11a}), to distinguish them from added planets in our simulations.  As shown in our results, the likelihood of all six confirmed planets to be transiting for our line-of-sight if the system contains a massive outer planet on an inclined orbit is bolstered considerably by their compact configuration that keeps the planets locked with low mutual inclinations over secular timescales, even if the six planets have a wide range of inclinations over the nodal precession cycle. So long as their mutual inclinations are very low, the probability to observe all six planets in transit is not significantly less than the transit probability of the outermost planet. If the planets had significant mutual inclinations, the likelihood that they would all be transiting for an observer from any favorable perspective is very small. The conditions under which an inclined Jovian companion breaks this coplanarity and makes it unlikely that all six could be transiting for any observer, provides upper limits on the mass and inclination of a Jovian planet to orbital distances beyond what is possible with existing TTV data and potentially beyond what is possible with existing RV data. We explain our method and results for this problem in the next section.

\section{TTV Constraints on Putative Planets}
The dataset of transit times for Kepler-11 modeled by \citet{liss13} included all short cadence \textit{Kepler} data available through Q14, with long cadence data where short cadence was unavailable. We complete the dataset using the long cadence transit times from Q15 through Q17 from \citet{rowe15a}. We fitted the transit times using the orbital periods, phases, eccentricity vector components and dynamical masses as free parameters for Kepler-11 b-f throughout. 

Kepler-11 g has a period more than 2.5 times that of with Kepler-11 f, and its mass is poorly constrained by the TTVs  \citep{liss13}. The planet is 3.3 R$_{\oplus}$ in size. In some cases, planetary radius can serve as a reasonable proxy for mass, although the diversity in density among sub-Neptunes, both between systems and within the same systems (\citealt{jont14,jont15,jont16}), leaves a wide range of plausible masses for Kepler-11 g.

For most of our simulations, we assumed the system was co-planar and fixed the mass of Kepler-11 g at 5.5 M$_{\oplus}$. Our estimate for the mass of this planet follows a simple approximate mass-radius relation for Kepler-11 b--f $\left( \frac{M_{p}}{M_{\oplus}} \sim \frac12 \left(\frac{R_p}{R_{\oplus}}\right)^{2}  \right)$. This is similar to the mass-radius relation for the Solar System found by \citet{liss11b} except that the planets of Kepler-11 are roughly half as massive over the same range in radii. This assumed mass for Kepler-11 g is close to the mode of well-characterized planets less massive than Neptune \citep{jont16}. 

\subsection{A putative planet between Kepler-11 f and Kepler-11 g.}
First, we tested the effect of an undetected planet orbiting between Kepler-11 f and Kepler-11 g on the goodness of fit of all measured transit times found via Levenberg-Marquardt minimization (L-M) for a range of orbital periods between Kepler-11 f and Kepler-11 g from 50 to 110 days. 

In our TTV modeling, for the inner five planets; orbital period, initial orbital phase, eccentricity vector components, and mass were all free parameters (5 per planet). We performed a grid search over a fixed orbital period for Kepler-11 x, with the orbital phase as a free parameter. In these fits, we fixed the eccentricity of Kepler-11 x at zero, under the assumption that since the TTVs in Kepler-11 g are sensitive to the relative eccentricities of Kepler-11 x and Kepler-11 g, freeing the eccentricity of either planet would have a similar effect but freeing both would add two parameters too many.

To map out the goodness-of-fit ($\chi^2$) as a function of the orbital period of Kepler-11 x,  we began this grid search near the middle of the orbital period range ($P_{x}$ =  80 days), using the best fit at each orbital period as an initial input model for the next orbital period, changing the fixed orbital period by 0.1 days between each L-M fit. We performed this search for three fixed masses for the undetected Kepler-11 x: 1 M$_{\oplus}$, 3 M$_{\oplus}$ and 10 M$_{\oplus}$. 

\begin{figure}[h!]
\includegraphics [height = 1.4 in]{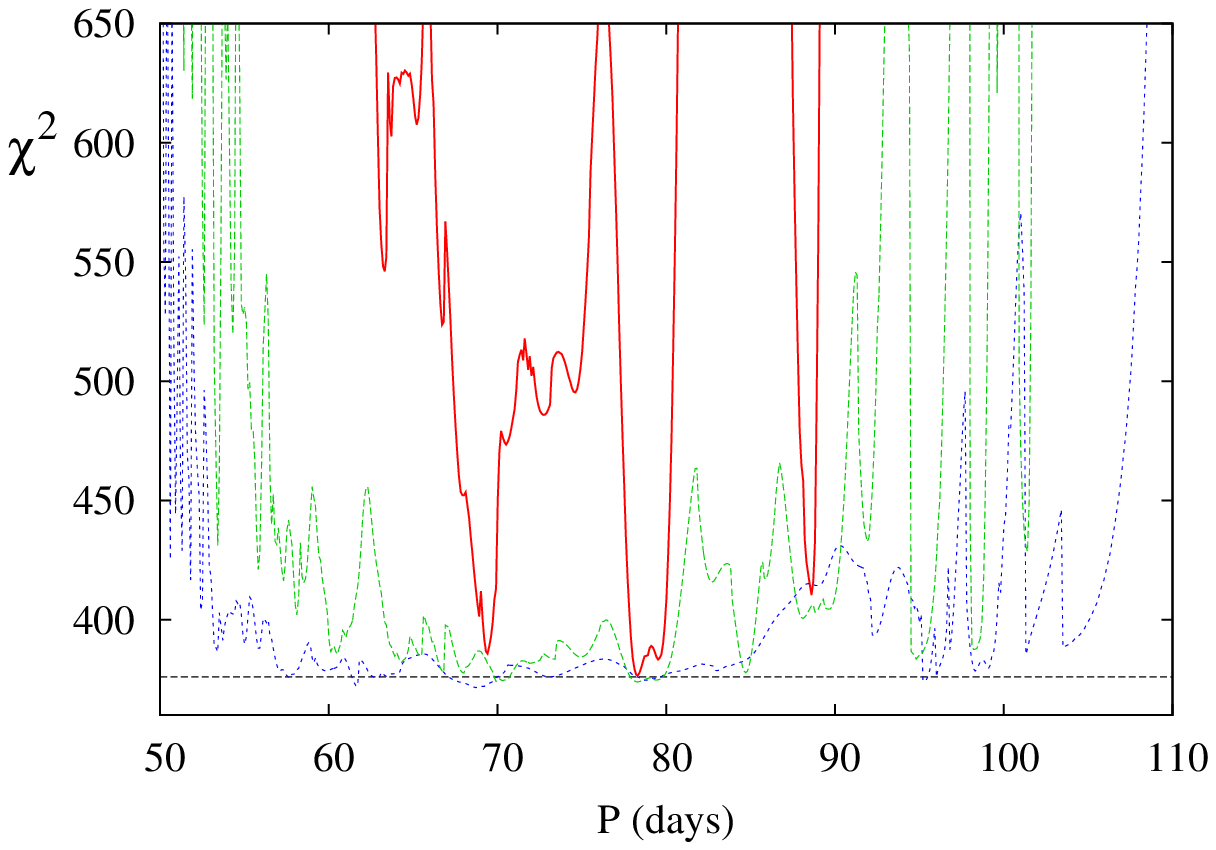}
\includegraphics [height = 1.4 in]{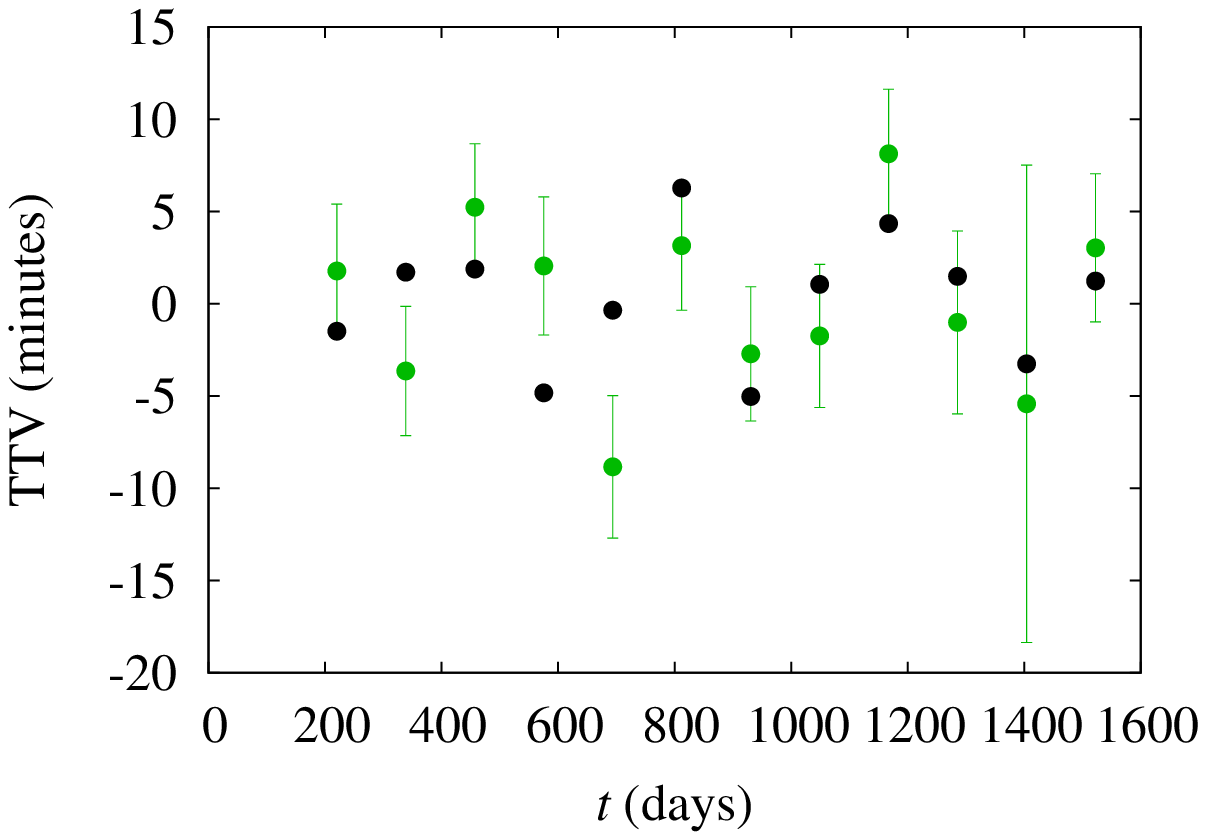}
\includegraphics [height = 1.4 in]{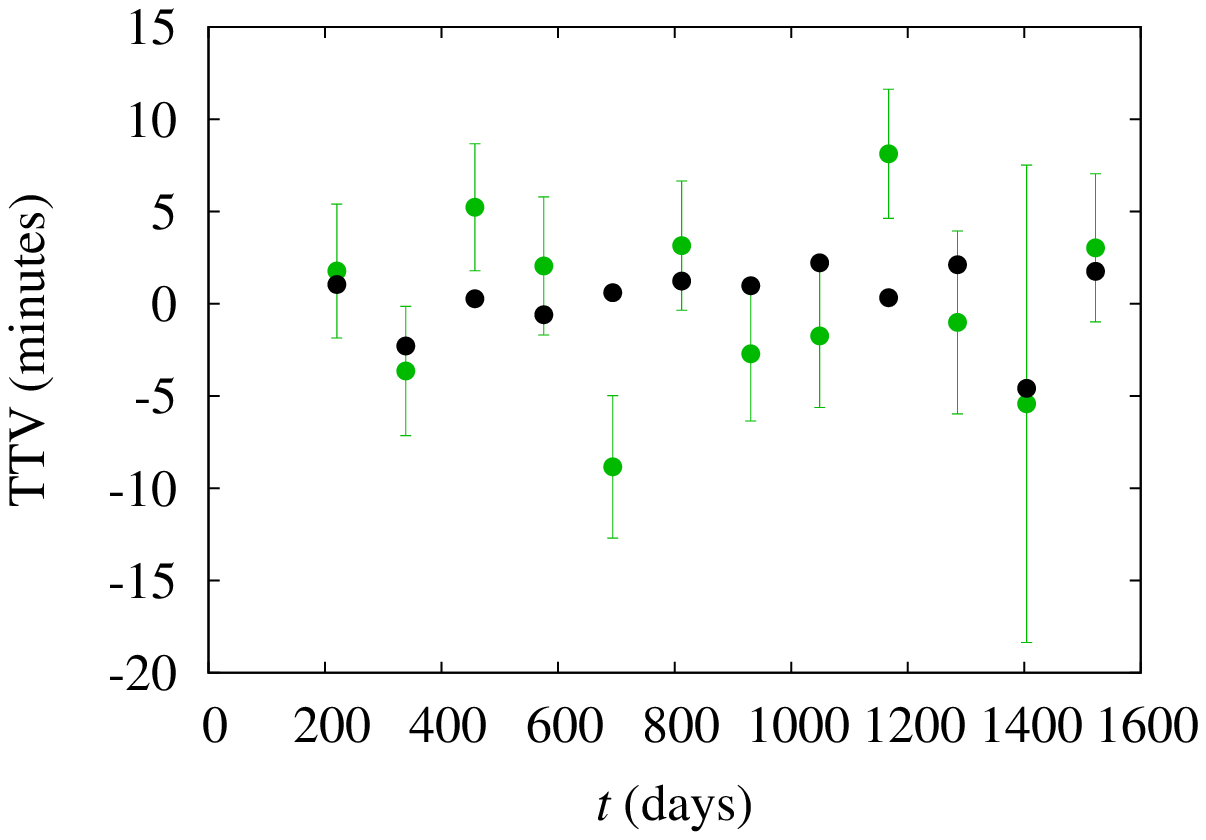}
\caption{The effect of undetected planets at Kepler-11 on transit times of Kepler-11 g. On the left, we plot $\chi^2$ for the best-fit TTV model at each period for a planet between Kepler-11 f and Kepler-11 g over a range of orbital periods for a 10 M$_{\oplus}$ (solid red curve), 3 M$_{\oplus}$ (dashed green curve) and 1 M$_{\oplus}$ (dotted blue curve) intermediate planet. In the case of a 3 M$_{\oplus}$ and 10 M$_{\oplus}$ planet the inclusion of this extra planet does not improve the TTV model over the 6 planet model where $\chi^2 = 374$, while the 1 M$_{\oplus}$ model planet gives a negligible improvement for a narrow range of orbital periods. The middle and right panels compare the observed TTVs (green) with theoretical TTVs (black) induced in Kepler-11 g from the best-fit 10 M$_{\oplus}$ planet orbiting at 69.2 days (left) and on the right the TTVs in Kepler-11 g from the best-fit 10 M$_{\oplus}$ planet orbiting at 78.1 days (right). Here the time is given as $t$ = BJD- 2,454,900.}
\label{fig:TTVsX} 
\end{figure}

We found that the goodness-of-fit ($\chi^2$) showed sharp features due to the sensitivity of multiple planets' transit times to a perturber between Kepler-11 f and Kepler-11 g. In most cases, we found that the extra parameters degraded rather than improved the goodness of fit (see left panel in Figure~\ref{fig:TTVsX}). The best fit model that we found with a Kepler-11 x had a $\chi^2$ value of 371, with the mass of Kepler-11 x fixed at 1 M$_{\oplus}$, and 4 additional free parameters, namely, the orbital period and phase of Kepler-11 x, and the eccentricity vector components of Kepler-11 g. The right two panels in Figure~\ref{fig:TTVsX} illustrate the effect of Kepler-11 x on the TTVs of Kepler-11 g, at two local minima found in the left panel. In both cases, high frequency TTVs (so-called synodic chopping) improves the goodness-of-fit at very specific orbital periods.

The Bayes Information Criterion offers a simplistic way to compare the best fit model with extra parameters to the best fit 6-planet model without Kepler-11 x as follows:
\begin{equation}
BIC = \chi^2+ k \ln{n}
\end{equation}
where $\chi^2$ measures of goodness-of-fit of a parametric model to the observed transit timing data, $k$ is the number of free parameters and $n$ is the number of measured transit times. In this case, the dataset includes the transit times of all 6 known planets (340 transits). For the 6-planet model, where $\chi^2_{6pl} = 374$, and $k_{6pl} = 27$  we found BIC$_{6pl}$ = 531. The corresponding estimate for our best fit model including Kepler-11 x gives BIC$_{x}$ of 552. Since the 6-planet model has a lower BIC than the model with Kepler-11 x, there is no significant evidence for the more complex model. 

In sum, we find that in most cases an additional planet between Kepler-11 f and Kepler-11 g degrades rather than enhances the fit to the TTV data, and where the fit is improved, the improvement provides no significant evidence of a planet between Kepler-11 f and Kepler-11 g. 

\subsection{A putative Jovian planet beyond Kepler-11 g.}
\begin{figure}[h!]
\includegraphics [height = 1.4 in]{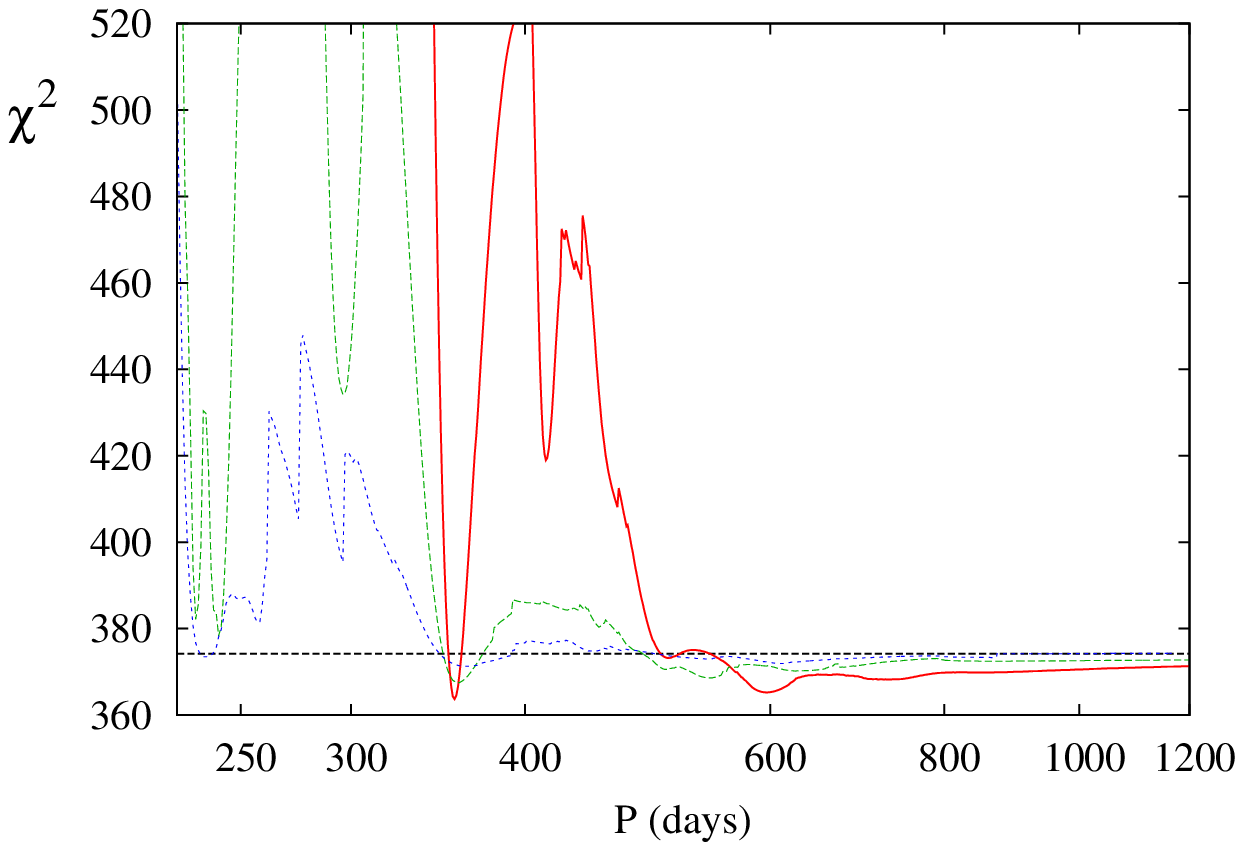}
\includegraphics [height = 1.4 in]{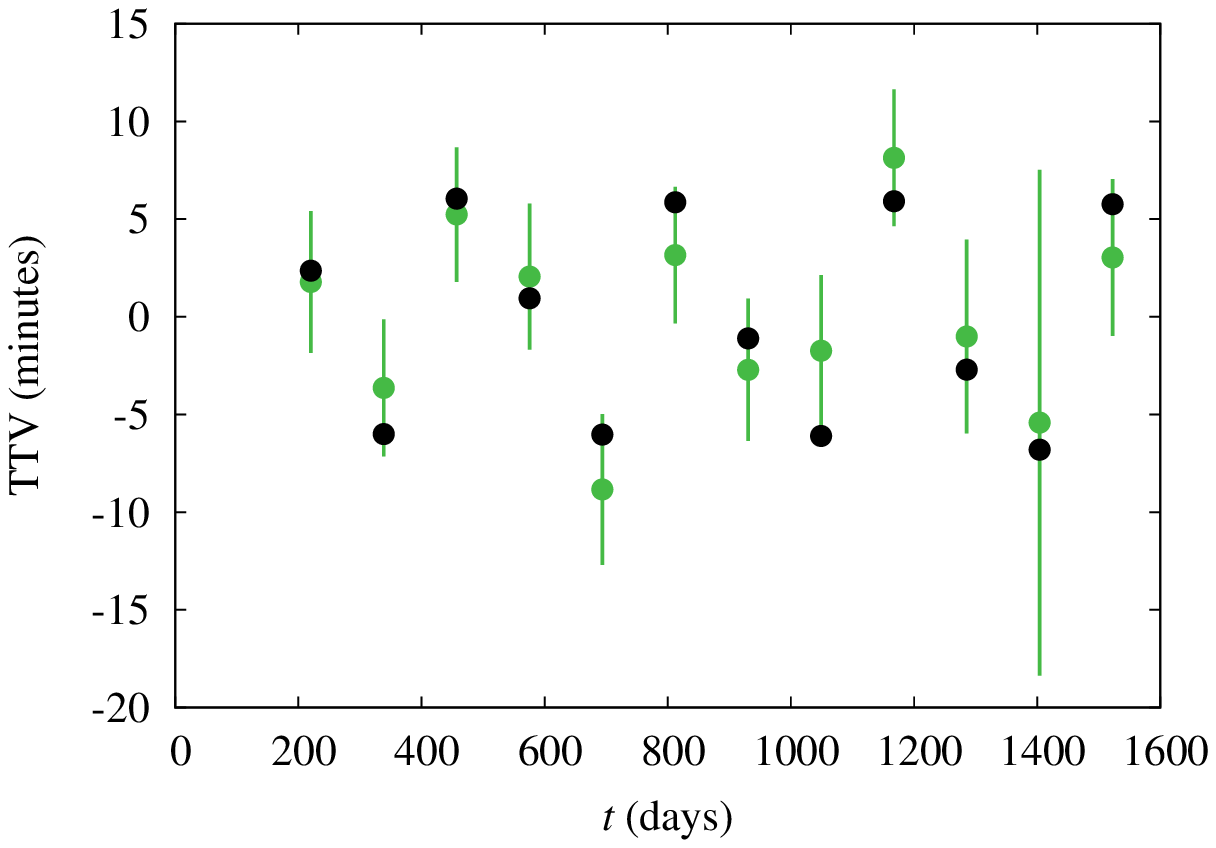}
\includegraphics [height = 1.4 in]{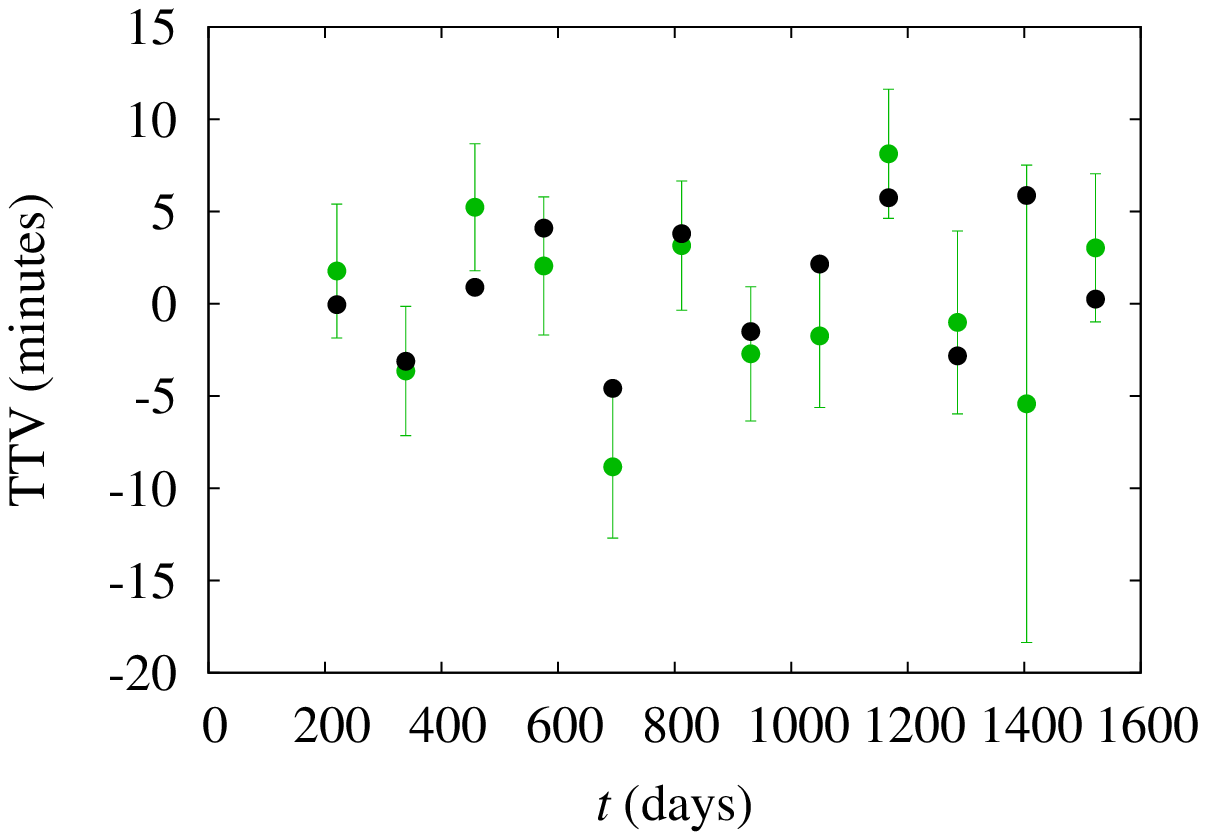}
\caption{The effect of a planet beyond Kepler-11 g on transit timing models. The left panel marks the $\chi^2$ of the best fit model at each possible orbital period for the putative perturbing planet: 1 M$_{Jup}$ planet (solid red curve), a 0.3 M$_{Jup}$ perturber (dashed green curve), and a 0.1 M$_{Jup}$ perturber (dotted blue curve). The middle and right panel show observed transit timing variations with uncertainties in green, and simulated transit times in black for best fit models with a Jovian-mass planet orbiting at 356 and 600 days respectively. }
\label{fig:TTVsJ} 
\end{figure}

We tested the effect of a Jovian-mass perturber (``Kepler-11 J") on an eccentric orbit beyond Kepler-11 g on the goodness-of-fit of the transit timing data for the six known planets. For these models, we fixed the orbital period of Kepler-11J for each L-M fit, increasing the orbital period in increments of 1 day from 250 to 1200 days, for three possible perturbing masses: 1 M$_{Jup}$, 0.3 M$_{Jup}$, and 0.1 M$_{Jup}$. To explore the goodness-of-fit we began our grid search at 800 days, using the local minimum found via L-M fitting for our initial parameters. After finding the local minima at 354 and 700 days, we repeated the grid search using these two locations as starting points. The eccentricity vector components, orbital period and initial orbital phase of the extra planet added four parameters to the TTV model. We show the resulting goodness-of-fit in the left panel of Figure~\ref{fig:TTVsJ}. At almost all distances, the extra parameters degraded the goodness-of-fit, implying that an extra planet out to some distance where its gravity has no material effect on the TTVs, is unlikely. We found two regions that slightly improved the fit to the observed TTVs by causing synodic chopping in the TTVs of Kepler-11 g, at orbital periods of 354 days and a broad region from 600--800 days respectively, as illustrated in the left panel of Figure~\ref{fig:TTVsJ}. We illustrate the TTVs of Kepler-11 g at both of these local minima in the right two panels of Figure~\ref{fig:TTVsJ}. \\

We compared the model of a Jovian planet at 354 days with the simpler model of no planets beyond Kepler-11 g (which has the same $\chi^2$ as a model with a Jovian planet beyond 1200 days.) The extra free parameters in the Jovian planet model include the orbital period, phase and eccentricity vector components of the Jovian (but its mass is fixed). Our best fit model included a Jovian planet at 354 days, with a goodness-of-fit $\chi^2_{J} = 364$ and $k_{J} = 31$, and hence BIC$_{J}$ = 545. 

We repeated this experiment with the mass of the perturber fixed at 0.3 M$_{Jup}$ and 0.1 M$_{Jup}$ respectively. Reducing the mass reduced the height and depth of local extrema substantially, whilst leaving the orbital periods of the best-fit solutions roughly unchanged.  

Since the 6-planet model has a lower BIC than the model with a Jovian planet, there is no significant evidence for a Jovian planet beyond Kepler-11 g from the TTVs. For the range of distances where the goodness-of-fit is substantially degraded, we note that a Jovian-mass planet is most unlikely within 480 days orbital period, with the exception of narrow possible range of orbital periods around 356 days. For a 0.3 M$_{Jup}$ perturber, a wider range of solutions are consistent with the data around 356 days orbital period, but otherwise this extra planet is unlikely within about 460 days (1.1 AU). \\

In the case of a 0.1 M$_{Jup}$ perturber at an orbital period of 241 days, we found a TTV model with $\chi^2$ = 373, comparable to the six-planet model. While such a planet is ruled unlikely over a much smaller range of distances from the TTVs (from about 250--350 days), the improvement to the TTV model around 356 days or further is also substantially weaker. Hence we conclude that the TTVs provide no strong evidence of an additional planet beyond Kepler-11 g, and a weak improvement in TTV fitting for a range of possible masses with a perturber orbiting around 354 days.  

We note that these lower mass solutions for the perturber may not be ruled out by the existing RV dataset. Hence we look for further dynamical constraints from outer perturbers and the extremely low mutual inclinations between the confirmed planets. 
\section{Coplanarity Constraints via Secular Inclination Evolution}
\subsection{Set-up}
Any two planets `y' and `z' have a mutual inclination defined by:
\begin{equation}
\cos\phi_{y,z} = \cos i_{y} \cos i_{z} + \sin i_{y} \sin i_{z} \cos(\Omega_{y}-\Omega_{z}),
\end{equation}
where $\phi_{y,z}$ is their true mutual inclinations, and $i$ and $\Omega$ are orbital inclinations and ascending nodes \citep{Ragozzine2010}.

We consider planet y to be ``nearly co-planar" with z (i.e., likely for both planets to transit for any observer that observes the outer planet to be transiting) if:
\begin{equation}
\sin\phi_{y,z}  \leq \frac{R_{\star}}{a_{z}},
\end{equation}
where $R_{\star}$ is the radius of the star and $a_{z}$ is the semi-major axis of the outer planet `z'. The actual likelihood depends on the impact parameter for the outer planet, ranging from 50\% if the outer planet has a grazing transit geometry, to 100\% if the outer planet has zero impact parameter for the observer.

For a simulation over the secular timescale we count what fraction of time-steps this condition is satisfied, with Kepler-11 g as the outer planet. Note, however, that this condition for coplanarity is commutative. Our results are unaffected by whether we count how many time-steps planet `b' (or `c', or `d') is in the plane of planet `g' or vice-versa. Although Kepler-11 b has a higher transit probability ($R_{\star}/a$) than its neighbors,  we choose the varying plane of Kepler-11 b as our reference plane because in practice, we find that the five planets Kepler-11 b--f have negligible mutual inclinations in all our simulations, and our results are driven by the separation of Kepler-11 g from the inner five. 

In our first set of simulations, we assume the system of transiting planets at Kepler-11 was initially coplanar. Although there are partial constraints on the present day relative inclinations from their modeled transit durations, these provide no information of the location of the ascending node ($\Omega$).  Without this crucial information, the observed transit impact parameters give approximate minimum mutual inclinations between the confirmed planets. We will test this configuration for our initial conditions, but for now assume the system was initially coplanar. This is the conservative choice for our investigation, since any initial non-zero mutual inclinations would likely reduce the fraction of the time that the planets are in a co-transiting configuration and increase the distance out to which we would find an inclined Jovian mass perturber to be unlikely. Therefore, by assuming an initial coplanar configuration we are measuring a minimum distance for an inclined perturber to be unlikely given that the six confirmed planets are co-transiting. 

Under this setup, without an inclined perturber, the confirmed transiting planets remain coplanar indefinitely. However, the addition of a hypothetical Jovian planet beyond Kepler-11 g (``Kepler-11 J") will cause oscillations in the inclination vector components of each transiting planet over secular timescales if it has some inclination relative to the compact transiting system. Whether or not these secular effects break the coplanarity of the system depends on Kepler-11 J's mass, orbital distance and inclination. The large gap in orbital period between Kepler-11 f and Kepler-11 g, combined with the compact configuration of the five inner planets, ensures that co-planarity is always lost when Kepler-11 g is excited to significant inclinations, while Kepler-11 b--f remain nearly coplanar. 

We considered a range of distances for the Jovian-mass perturber from 1.5 to 6 AU, with initial inclinations at 1, 3 and 10 degrees, and determined which combinations of inclination and orbital distance would cause the co-transiting configuration of Kepler-11 b--g to be ``unlikely".  We chose this range of mutual inclinations to explore for the Kepler-11 system to be consistent with the moderate inclinations observed in multi-planet systems, including the Solar System. We note that in the Solar System, Jupiter has a mutual inclination of 2$^{\circ}$ with Venus, and 6$^{\circ}$ with Mercury. Furthermore, the normalized transit duration ratios of all of \textit{Kepler}'s multiplanet systems are consistent with typical mutual inclinations of 1.0$^{\circ}$--2.0$^{\circ}$ (\citealt{fab14}). 

For all simulations, we adopt the planetary and stellar parameters of Kepler-11 in Tables 3 and 4 of \citet{liss13}. These assumed perfectly co-planar orbits. Using transit timing data through Q14 of the \textit{Kepler} mission, \citet{liss13} found tight constraints on the masses and orbital eccentricities of Kepler-11 b-f, and measured the mass of Kepler-11 M$_{\star}$ = 0.961 M$_{\odot}$. In these simulations we assumed a mass of 5.5$M_{\oplus}$ for Kepler-11 g, as explained in $\S$2.2. This assumed mass is based on a simple mass-radius relation for the well-characterized planets at Kepler-11 and the mode of well-characterized low-mass exoplanet. However, it is roughly half of the mean mass for planets of its size, based on the probabilistic planetary mass-radius relationship inferred from follow-up of Kepler planet candidates (\citealt{Wolfgang2016,Chen2017}) and we tested our results with a mass of 10 M$_{\oplus}$ for Kepler-11 g.  

We used a hybrid symplectic integrator from the Mercury package \citep{Chambers1999}, with a time-step set to 0.1 days, roughly 0.01 times the orbital period of the innermost planet in our simulations.
This choice of time-step optimized numerical accuracy and efficiency, since shorter time-steps provided no changes in the output (to ten significant figures) and hence no improvement in accuracy. 

In our models with an initial inclination for Kepler-11J at 3$^\circ$, we found that if it is further than 3.0 AU, the cotransiting configuration of Kepler-11 b--g is likely. We set this model of Kepler-11 J as our benchmark for comparison as we considered additional effects on the coplanarity of Kepler-11 b--g, beginning with a test on how sensitive the coplanarity of the Kepler-11 system is to the chosen mass of Kepler-11 g (see $\S$3.2.1). 

We also considered the possibility of a seventh low-mass coplanar planet, ``Kepler-11 x", orbiting between Kepler-11 f and Kepler-11 g, and its effect on the coplanarity of the known transiting planets with a nearby Jovian planet on an inclined orbit (s3.2.2). Such a planet has not been detected in the \textit{Kepler} light curve, indicating that it is either: i) too small to cause a detectable transit, ii) inclined relative to the six known planets, or iii) non-existent. Nevertheless, if it were to exist and if it were nearly coplanar with the known transiting planets, it could strengthen the coplanar bond between the known planets and weaken the constraints of the inclination, mass and minimum orbital distance of Kepler-11 J. We ran three suites of models with the low mass planet between Kepler-11 f and Kepler-11 g set at  2.4$\times 10^{-6}$ M$_{\star}$ ($\approx$1 M$_{\oplus}$) and $\approx$2 M$_{\oplus}$ and $\approx$3 M$_{\oplus}$ respectively, and considered a range of distances from 0.29 to 0.43 AU. This range was chosen to include all distances between Kepler-11 f and Kepler-11 g for which the system would remain stable for 8 Myr. Note that the result of our TTV studies (Figure~\ref{fig:TTVsX}) suggest that a 3 $M_{\oplus}$ or a 10 $M_{\oplus}$ planet in this region is unlikely for most of the range of possible orbital periods between 60 and 100 days and we see little or no evidence for (or against) a 1 $M_{\oplus}$ planet.

As an additional test on our nominal result, we relaxed our assumption of initially coplanar orbits for the transiting planets ($\S$3.2.4). \citet{liss13} show that there is a minimum mutual inclination between the planets given the constraints on their transit impact parameters. While substantial mutual inclinations between the planets are possible, this is exceedingly unlikely, since it would require the planets to share a common longitude of ascending node with our line of sight. Hence, the impact parameters give an approximate upper limit to the mutual inclinations of the planets. For our final test, we adopt the nominal lower bounds on inclinations of the \citet{liss13} to set initial mutual inclinations between Kepler-11 b and all of its neighbors.

Finally, since the angular momentum of the system is dominated by the distant perturber, we tested the effect of reducing its the mass on whether the unlikely co-transiting state of a Kepler-11 b--g persists with a lower mass perturber inclined at 3$^{\circ}$ ($\S$3.2.3). 

\begin{table}[!h]
\tiny
    \centering
     \begin{tabular}{|l||c|c|c|c|c||}
     \hline  &  \textbf{Std. Trials} & \textbf{m$_g$ = 10 M$_{\oplus}$} & \textbf{ m$_x$ = 1,2,3 M$_{\oplus}$} &  \textbf{ Initial Inc. Kep 11 b--g} & \textbf{m$_{J}$ = 0.1, 0.3 M$_{Jup}$}   \\
       \hline
      \textbf{Inc. (J)}  & 1$^{\circ}$,3$^{\circ}$,10$^{\circ}$  & 3$^{\circ}$ & 3$^{\circ}$  & 3$^{\circ}$ & 3$^{\circ}$ \\
 \hline \textbf{`a'  (AU)} & a$_{J}$: 1.5--6.0 & a$_{J}$: 1.5-3.5 & a$_{x}$: 0.29-0.43 & a$_{J}$: 1.5--6.0 &  a$_{J}$: 0.8-3.0   \\ 
 \hline Figure & 4 & 5 & 6 & 7 & 9 \\
 \hline
    \end{tabular}
    \caption{Table of the initial conditions in each set of simulations. Our standard trials include the six confirmed transiting planets of Kepler-11 on co-planar orbits with their masses and other orbital parameters fixed at the best values published by \citet{liss13}, M$_{J} = 1$ M$_{Jup}$, and the mass of Kepler-11 g fixed at 5.5 M$_{\oplus}$. 
    The next column describes simulations with a 10 M$_{\oplus}$ mass for Kepler-11 g. The subsequent column describes simulations with the same parameters as our standard trial but including a 1, 2 or 3 M$_{\oplus}$ non-transiting planet between Kepler-11 f and Kepler-11 g (``Kepler-11 x").  The next column includes simulations where the initial mutual inclinations between the transiting planets are non-zero, based on the results of \citet{liss13}. The final column summarizes the simulations that we performed with smaller masses for Kepler-11 J (at 0.1 and 0.3 M$_{Jup}$).  
    The middle row indicates the range in initial orbital distances explored for each set. In each case, we explore a range of distances and inclinations for Kepler-11 J that includes where the co-transiting probability transitions from $> 50 \%$ to $<50 \%$. The bottom row indicates the figure where the results are plotted.}
    \label{tbl:sims}
\end{table}

We summarize the input parameters all our simulations in Table~\ref{tbl:sims}.

\subsection{Results}
\begin{figure}[h!]
\includegraphics [height = 3.3 in, angle = 270 ]{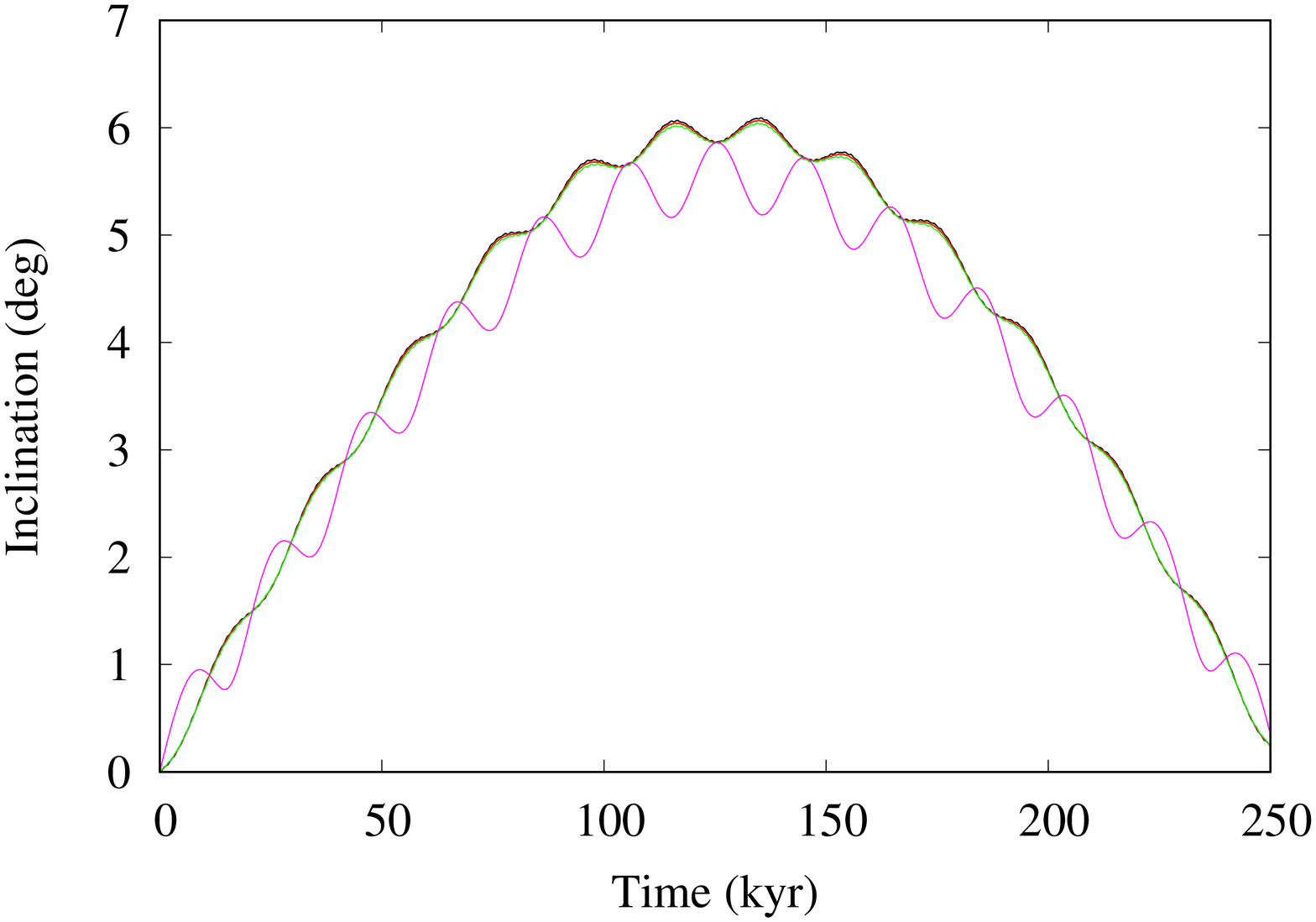}
\includegraphics [height = 3.3 in, angle = 270 ]{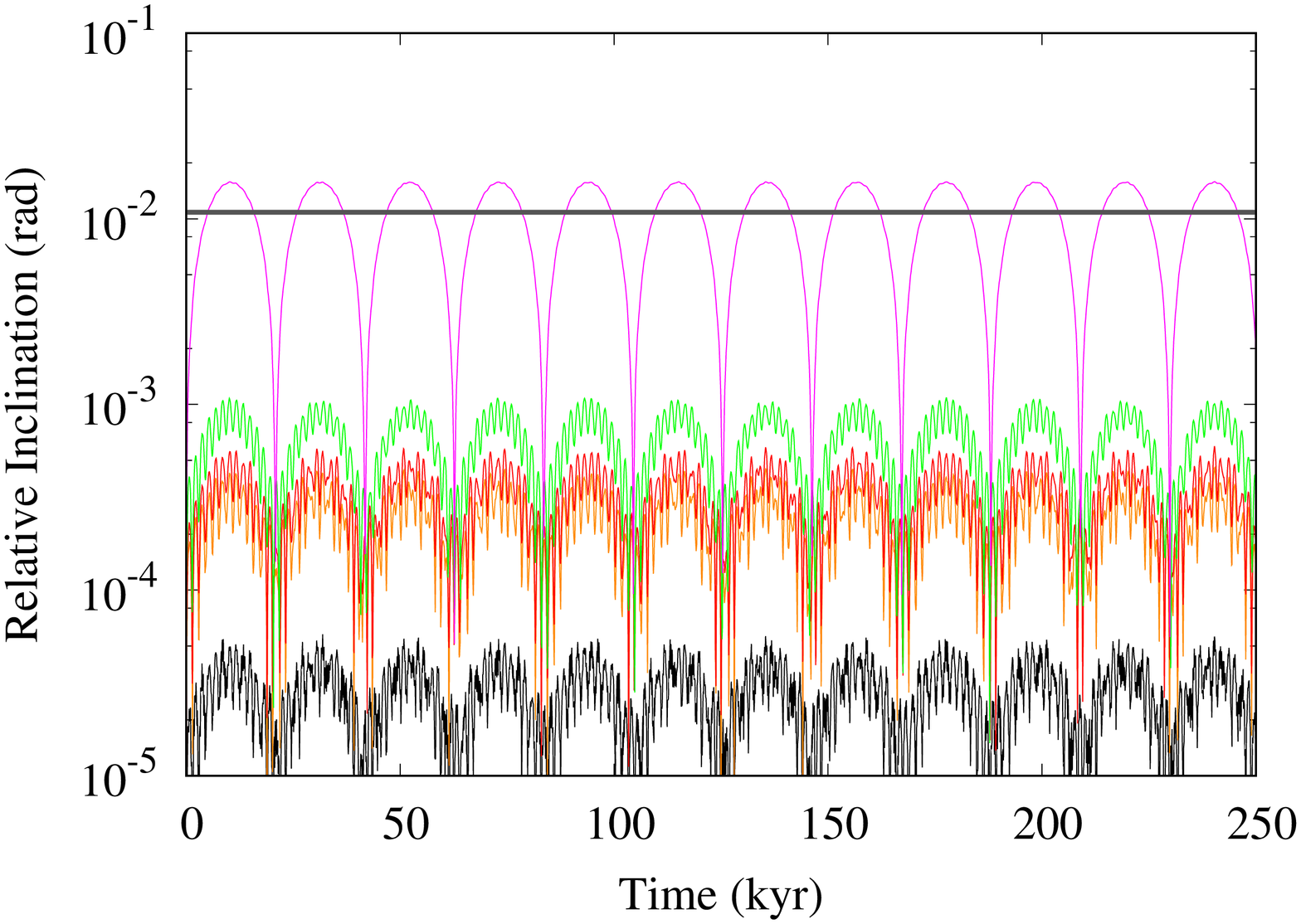}
\caption{Inclinations and relative inclinations for the six confirmed planets orbiting Kepler-11 (over the first 250 kyr of an 8 Myr simulation), due to perturbations of a 1 M$_{Jup}$ planet at 3.0 AU with an initial 3 degree inclination. The left panel shows the inclinations of all six transiting planets over time relative to their initial orbital plane (in degrees). The right panel shows inclinations relative to the contemporaneous orbital plane Kepler-11 b (logarithmic scale, in radians). The maximum relative inclination increases with increasing orbital distance from Kepler-11 b, with Kepler-11 c, d, e, f and g in black, orange, red, green and magenta respectively. The grey horizontal line marks $(R_{\star}/a)_{Kepler-11 g}$. When all planets are below the grey line, they are considered co-planar or `likely co-transiting'. When Kepler-11 g is above the line, they are unlikely to be co-transiting for an observer that observes the inner five to be co-transiting. In this configuration, they are co-transiting just over 50\% of the time, and hence the presence of Jovian planet at 3.0 AU or further cannot be ruled unlikely.}
\label{fig:coplanar} 
\end{figure}
We illustrate the co-planarity of the six known planets in Kepler-11 over 100 kyr in Figure~\ref{fig:coplanar}. The inner five planets of Kepler-11 are tightly coupled over secular timescales, with mutual inclinations never exceeding a few arc-minutes, while Kepler-11 g reaches an inclination of 1.4$^{\circ}$ during the secular cycle. The decoupling of Kepler-11 g from its inner neighbors makes it non-transiting for most observers that detect Kepler-11 b--f to be transiting for much of the time.  

We compared the numerical findings in Fig.~\ref{fig:coplanar} with the analytical solutions of \citet{Lai2016}. The coupling between Kepler-11 f and g at their respective semi-major axes ($a_f$ and $a_g$), when perturbed by Kepler-11 J is determined by the parameter 
\begin{equation}
\epsilon_{fg} = \hat{\Omega}_{fJ}  \frac{(\Omega_{gJ}/\Omega_{fJ})-1}{1+(L_{f}/L_{g})},
\end{equation}
where $L_{f}/L_{g} = (m_{f}/m_{g})(a_{f}/a_{g})^{1/2}$ is the ratio of the planets' angular momenta, $\Omega_{gJ}/\Omega_{fJ} \approx \left(\frac{a_g}{a_f}\right)^{3/2}$ is the ratio of precession rates of Kepler-11 f and Kepler-11 g respectively, driven by Kepler-11 J:
\begin{equation}
\Omega_{fJ} = \frac{Gm_{f}m_{g}a_{f}}{4a_{g}^2L_{1}}b_{3/2}^{(1)}\left(\frac{a_{f}}{a_{g}} \right). 
\end{equation}
Here, $b_{3/2}^{(1)}\left(\frac{a_{f}}{a_{g}}\right)$ is a Laplace coefficient in the literal expansion of the disturbing function \citep{Murray1999}. A similar expression exists for $\Omega_{gJ}$.  The measure of coupling $\epsilon$ determines how much inclination dispersion can be expected between the inner planets. Where $\epsilon \ll 1$, the inner planets are tightly coupled and remain coplanar, and where $\epsilon \gg 1$ mutual inclinations can reach as high as twice the inclination of the distant perturber \citep{Lai2016}.

Treating the system as two inner planets (Kepler-11 f and g) with a Jovian mass perturber at 3.0 AU inclined at 3$^{\circ}$, Kepler-11 f and Kepler-11 g are moderately coupled ($\epsilon = 0.2$), while the inner five are all strongly coupled such that between any two adjacent neighbors, $\epsilon \lesssim 0.001$. Treating the inner six as a group, and treating Kepler-11 e (8.0 M$_{\oplus}$) as the dominant mass within the group, the averaged coupling parameter $\hat{\epsilon} = 0.1$ and the spread in mutual inclinations for the inner six is 0.5$^{\circ}$, in close agreement with range in mutual inclinations shown in the left panel of Figure~\ref{fig:coplanar}. 

\begin{figure}[h!]
\includegraphics [height = 4.5 in, angle = 270 ]{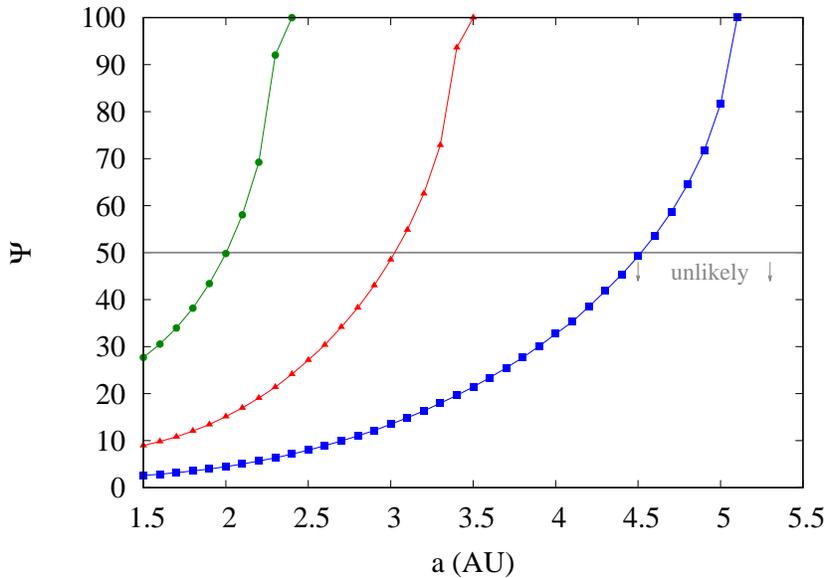}
\caption{Fraction of time all of Kepler-11 b--g are co-transiting ($\Psi$, given as a percentage). Simulations with a 1 M$_{Jup}$ planet at the distances indicated on the horizontal axis, and inclined at 1$^{\circ}$ are marked with green circles, at 3$^{\circ}$ by red triangles and by 10$^{\circ} $ by blue squares. Below the 50$\%$ line, the observed configuration of transiting planets is unlikely, while above the line it is plausible for any observers that observe Kepler-11 g to be transiting.}
\label{fig:fractransiting} 
\end{figure}

For our numerical determination of the likelihood of the inner six to be co-transiting, we counted the fraction of output time-steps over an 8 Myr simulation that Kepler-11 b-g are all likely transiting and show the result in Figure~\ref{fig:fractransiting}. The 8 Myr timescale covers $\gtrsim$20 secular periods for the inner planets, so that the fraction of time when the planets are co-planar is adequately sampled with at least one hundred time samples within each secular cycle. 

Figure~\ref{fig:fractransiting} highlights the sensitivity of the transiting configuration to a Jovian planet placed at different distances for three possible initial inclinations. For the nominal case of a 3$^{\circ}$ inclination, the co-transiting configuration becomes unlikely if the Jovian-mass planet is within 3.0 AU. With an initial inclination at 10$^{\circ}$, a Jovian-mass planet is unlikely interior to 4.5 AU. Even an inclination of just 1$^{\circ}$ provides an important constraint on a Jovian-mass perturber. Such a planet is unlikely to exist within 2.0 AU; comparable to the known constraint from RV spectroscopy.

We tested the sensitivity of our nominal result to the mass of Kepler-11 g, the presence of a planet orbiting between Kepler-11 f and Kepler-11 g, the mass of the perturbing outer planet, and the minimum mutual inclinations among the transiting planets derived from the transit light curve. 

\subsubsection{Sensitivity to the poorly constrained mass of Kepler-11 g.}

\begin{figure}[h!]
\includegraphics [height = 4.5 in, angle = 270 ]{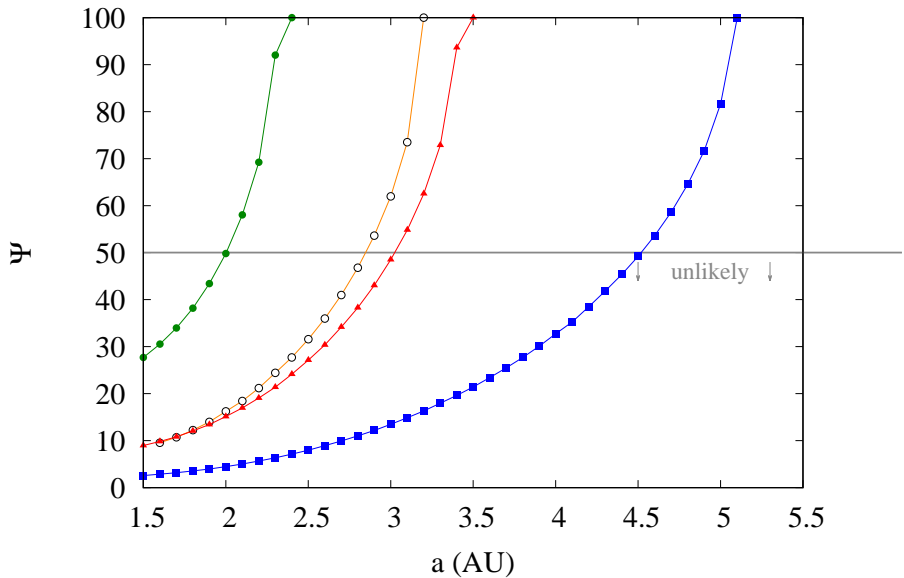}
\caption{The effect of the mass of Kepler-11 g on the fraction of time Kepler-11 b--g are co-transiting ($\Psi$, given as a percentage). We include our nominal models where the mass of Kepler-11 g is 5.5 $M_{\oplus}$, for three possible inclinations of the inclined perturber Kepler-11 J (1$^{\circ}$, green circles; 3$^{\circ}$, red triangles; and 10$^{\circ}$, blue squares). The open circles mark the fraction of the time that Kepler-11 b--g are co-transiting where Kepler-11 g has a mass of 10 M$_{\oplus}$, and the Jovian planet has an initial inclination of $3^{\circ}$.}
\label{fig:massG} 
\end{figure}

Figure~\ref{fig:massG} shows the sensitivity of our benchmark result to the mass of Kepler-11 g, which is poorly constrained from observations (Lissauer et al. 2013).  We find that increasing the mass of Kepler-11 g increases the fraction of the time that the 6 transiting planets are kept in a coplanar configuration by a few percent if Kepler-11J orbits between 2 and 3 AU from the star; i.e., increasing the mass of Kepler-11 g  by a factor of 2 modestly reduces the distance out to which we can rule that a Jovian perturber on a 3 degree inclination is unlikely from 3.0 to 2.8 AU. This is a far weaker effect than the initial inclination of the perturbing outer planet, and hence our results are insensitive to the mass of Kepler-11 g. 

\subsubsection{Senstivity to a hypothetical planet orbiting between Kepler-11 f and Kepler-11 g.}

\begin{figure}[h!]
\includegraphics [height = 4.5 in, angle = 270 ]{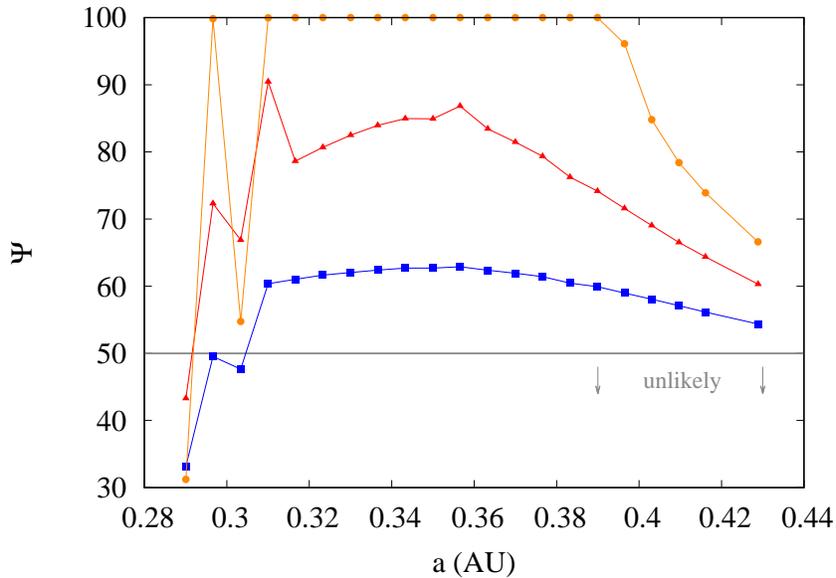}
\caption{The percentage of the time $\Psi$ that the transiting planets are ``coplanar" over an 8 Myr period, with an undetected coplanar planet orbiting between Kepler 11 f and Kepler-11 g. In the absence of such a planet, the likelihood that the known planets are co-transiting in this case is 50\%, assuming a Jovian mass planet inclined 3$^{\circ}$ and orbiting at 3.0 AU. We mark the enhanced likelihood of coplanarity with a 1 M$_{\oplus}$ planet (blue squares), a 2 M$_{\oplus}$ planet (orange circles), and a 3 M$_{\oplus}$ planet (red triangles), over a range of orbital distances between Kepler-11 f and Kepler-11 g. 
\label{fig:fractransitingX} }
\end{figure}

Figure~\ref{fig:fractransitingX} shows the effects of an undetected planet between Kepler-11 f and Kepler-11 g (with its mass at 5.5 M$_{\oplus}$) on the likelihood that the known planets are co-transiting. This would cause the orbital plane of the transiting planets to be more tightly coupled, requiring a perturbing Jovian planet to be either closer or more highly inclined to break the co-transiting configuration, and hence weakening our constraints. For these models, we assumed a Jovian mass planet (``Kepler-11 J") on a 3$^{\circ}$ inclination orbiting at 3.0 AU. Giving Kepler-11 x a mass of 1 M$_{\oplus}$ enhances the likelihood of the observed planets to be co-transiting, although fairly moderately. The likelihood of all planets transiting remains between 50\% and 63\% over the entire range of orbital distances where the system was stable for at least 8 Myr. 

Increasing the mass of Kepler-11 x to 2 or 3 M$_{\oplus}$ raises the likelihood that the known planets remain co-planar significantly. From figure~\ref{fig:fractransitingX}, it is clear that the coplanar condition is sensitive to the mass of the intermediate planet. In the case of the 3 M$_{\oplus}$ model planet, the 6 confirmed planets are co-transiting  all of the time for almost the entire range of possible orbital distances for an intermediate planet, with the exception of a the feature near 0.3 AU, near the 4:3 resonance with Kepler-11 f and the 2:1 resonance with Kepler-11 g. 

A 3 M$_{\oplus}$ intermediate planet Kepler-11 x would reduce the distance out to which a Jovian-mass planet can be ruled out on the condition of co-planarity, but a 3 M$_{\oplus}$ planet between Kepler-11 f and Kepler-11 g is also disfavored with the transit timing data as shown in section 2. A 2 M$_{\oplus}$ planet permits enough mutual inclination between Kepler-11 g and the inner five to make the cotransiting configuration likely but not 100\% of the time. A planet less massive than 1 M$_{\oplus}$ between Kepler-11 f and Kepler-11 g has an even weaker effect on the coplanarity of the system. We conclude here that while the TTVs disfavor a putative Kepler-11 x more massive than 3 M$_{\oplus}$, neither TTVs nor the constraint of coplanarity rule out a lower mass planet between Kepler-11 f and Kepler-11 g.

\subsubsection{Senstivity to initial inclinations of the transiting planets.}
Here, we return to our assumption that an initially perfectly co-planar configuration for the transiting planets gives a conservative minimum distance out to which a Jovian perturber is unlikely. We repeated our nominal simulation with Kepler-11 J at 3.0 AU with a inclination of 3$^{\circ}$ with respect to the initial orbit plane of Kepler-11 b, and with the initial relative inclinations of the transiting planets based on the values listed in Table 2 of \citet{liss13}. This assumes the planets have the same ascending node. The relative inclinations of the transiting planets at Kepler-11 to Kepler-11 b are all less than 0.2$^{\circ}$ except Kepler-11 e, which is mutually inclined to the orbit of Kepler-11 b by 0.75$^{\circ}$ under this assumption. 

\begin{figure}[h!]
\includegraphics [height = 2.1 in, angle = 270 ]{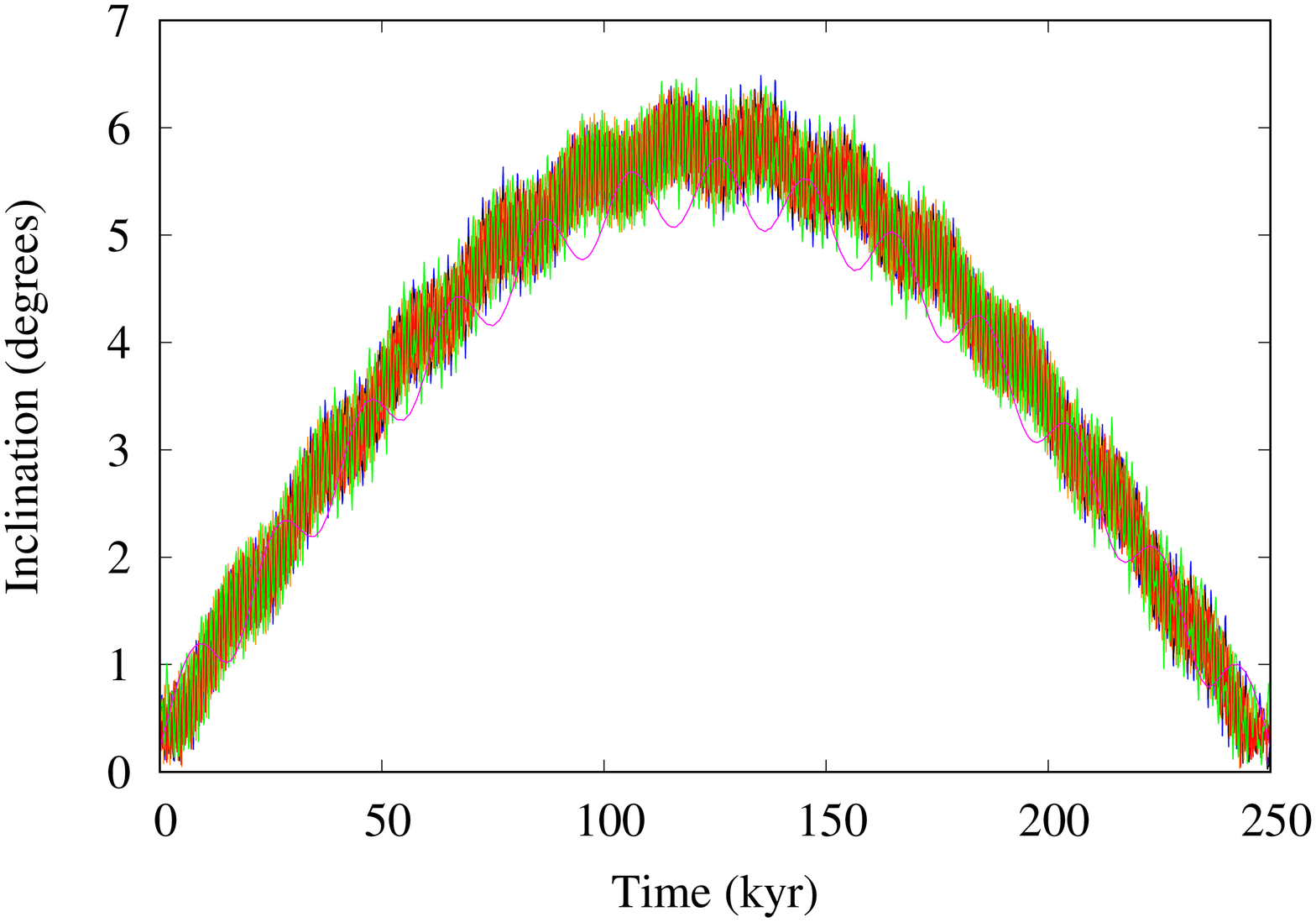}
\includegraphics [height = 2.1 in, angle = 270 ]{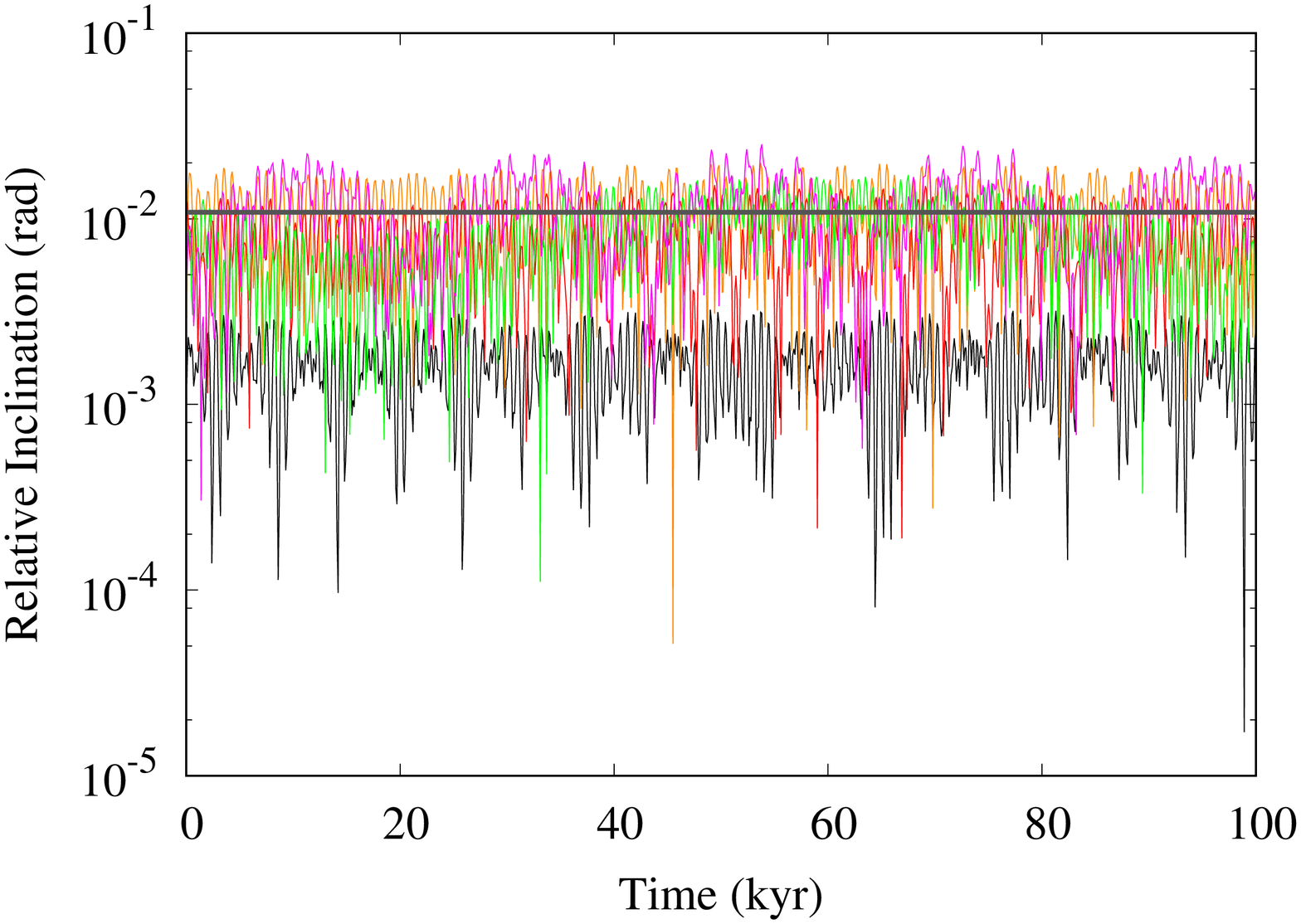}
\includegraphics [height = 2.1 in, angle = 270 ]{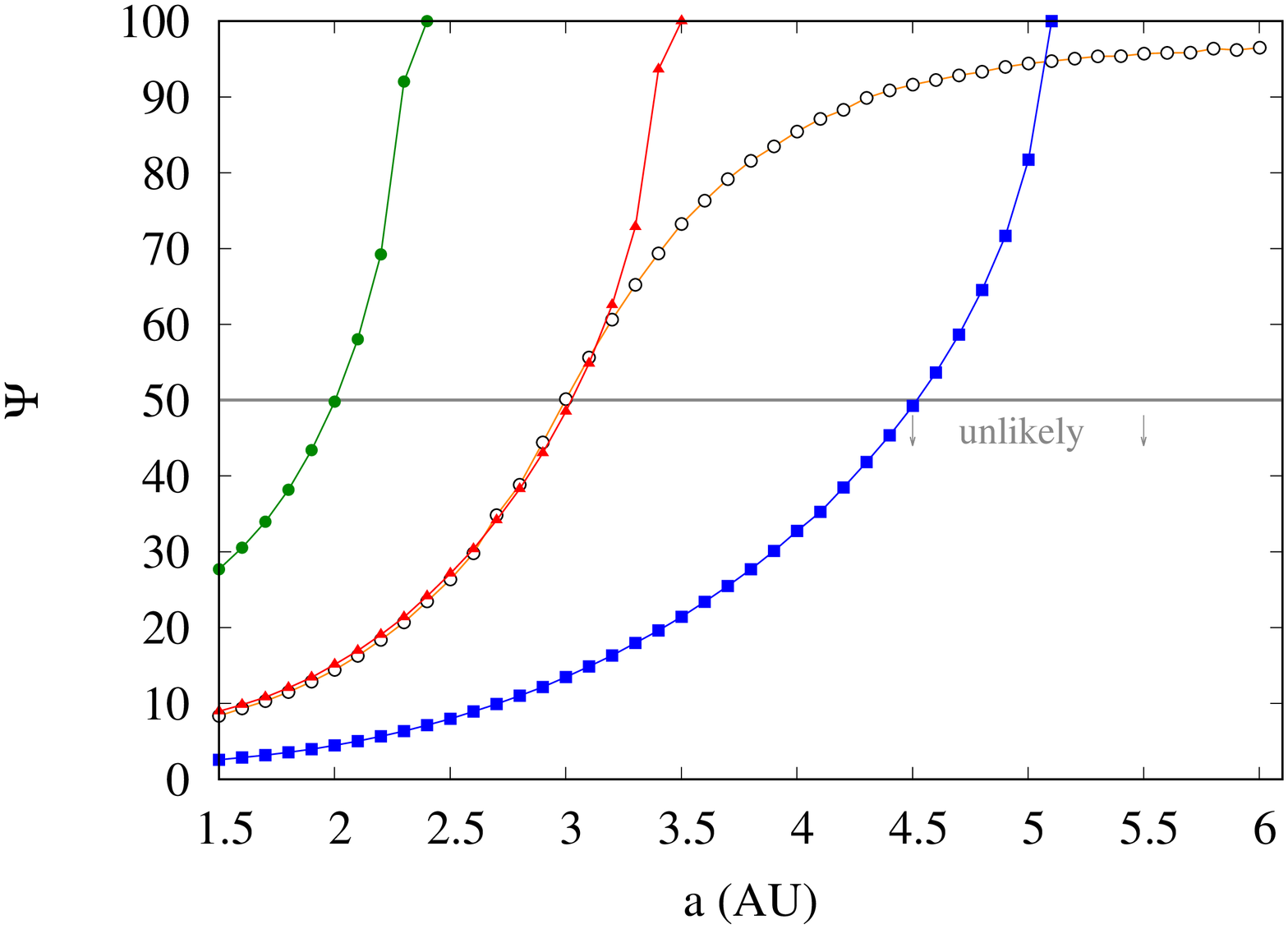}
\caption{The effect of a M$_{Jup}$ perturber with an initial inclination of 3$^{\circ}$, with initial mutual inclinations for the transiting planets set to their nominal lower bounds from the light curve modeling \citep{liss13}. The top left panel marks the inclination variations of the six transiting planets, with relative inclinations to Kepler-11 b marks in the second panel with logarithmic scale, with colors corresponding to the curves in Figure~\ref{fig:coplanar}. The top right panel shows that relative inclination peaks are not driven solely by Kepler-11 g (magenta), but also by Kepler-11 e (which has the highest initial inclination, shown in orange here), and occasionally Kepler-11 f (green). The third panel compares the effect of the distance of Kepler-11J on the fraction of the time that planets are co-transiting, with our nominal models with zero initial mutual inclinations between the transiting planets. Differences only arise if Kepler-11J is beyond 3.0 AU, where neither model can be ruled out with confidence.}
\label{fig:InitInc} 
\end{figure}

The results of our simulations with these initial mutual inclinations are shown in Figure~\ref{fig:InitInc}. We found that the range of mutual inclinations between Kepler-11 b--f is significantly higher than the initially co-planar model, and the constraints on a Jovian perturber are driven by relative inclination peaks in Kepler-11 e, f and g. The bottom panel of Figure~\ref{fig:InitInc} reveals important behavior that is different when initial mutual inclinations are included. With the Jovian planet at greater distances, instead of the likelihood of the co-transiting increasing rapidly to 100\%, the likelihood asymptotes to 100\% since with non-zero initial inclinations, the system is more sensitive to distant perturbations. For a Jovian perturber beyond 3.0 AU, our constraints from an initially co-planar system are more conservative, as we expected from our initial assumption. However, where there is a transition from ``unlikely" to ``more likely than not" (the  50\% line crossed near 3.0 AU), the initial mutual inclinations make little difference, since the range in mutual inclinations between the inner five and the weak coupling of Kepler-11 g have a comparable effect on the overall range in mutual inclinations.  \\

\subsubsection{Senstivity to the mass of the perturbing outer planet.}

\begin{figure}[h!]
\includegraphics [height = 3.3 in, angle = 270 ]{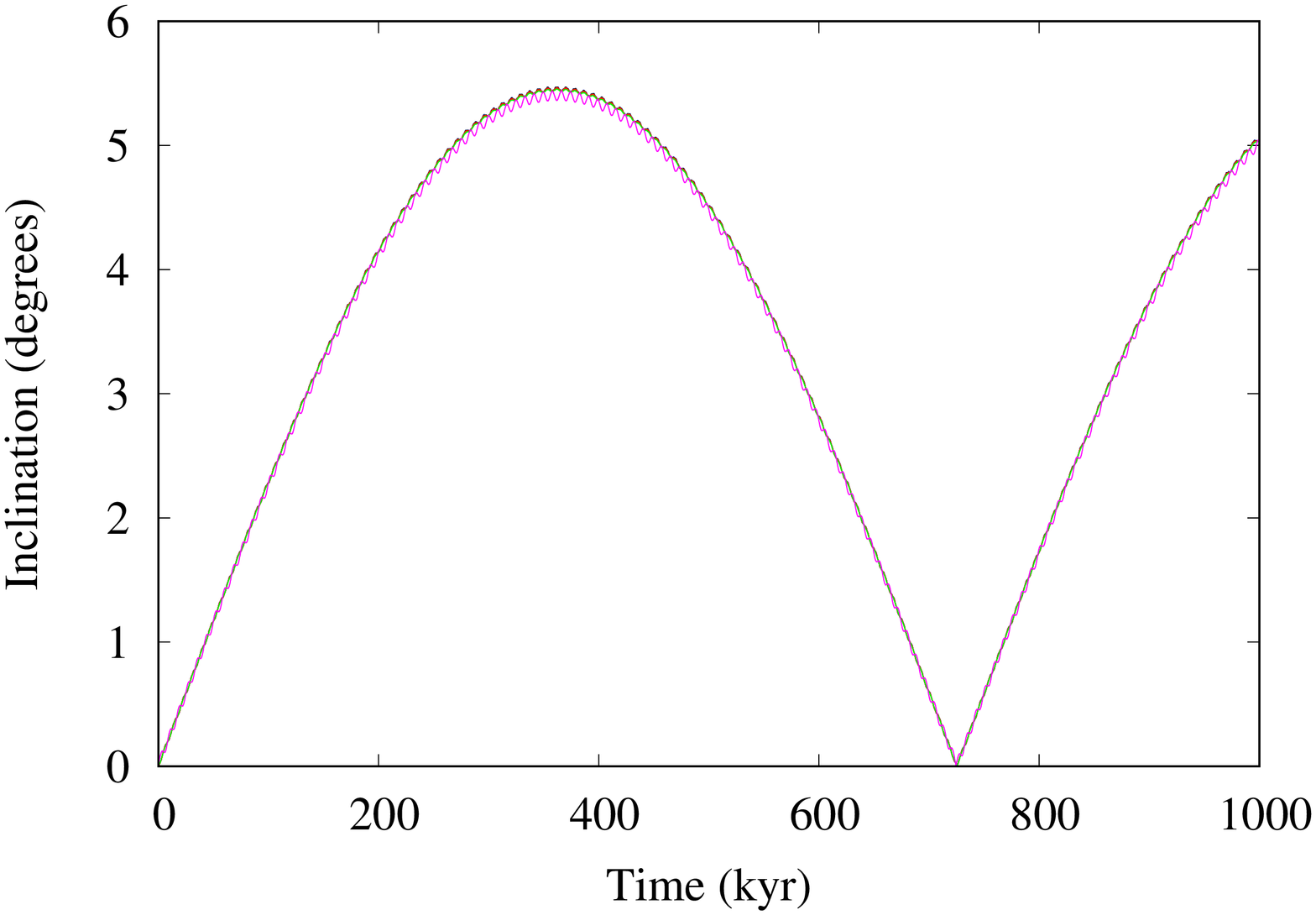}
\includegraphics [height = 3.3 in, angle = 270 ]{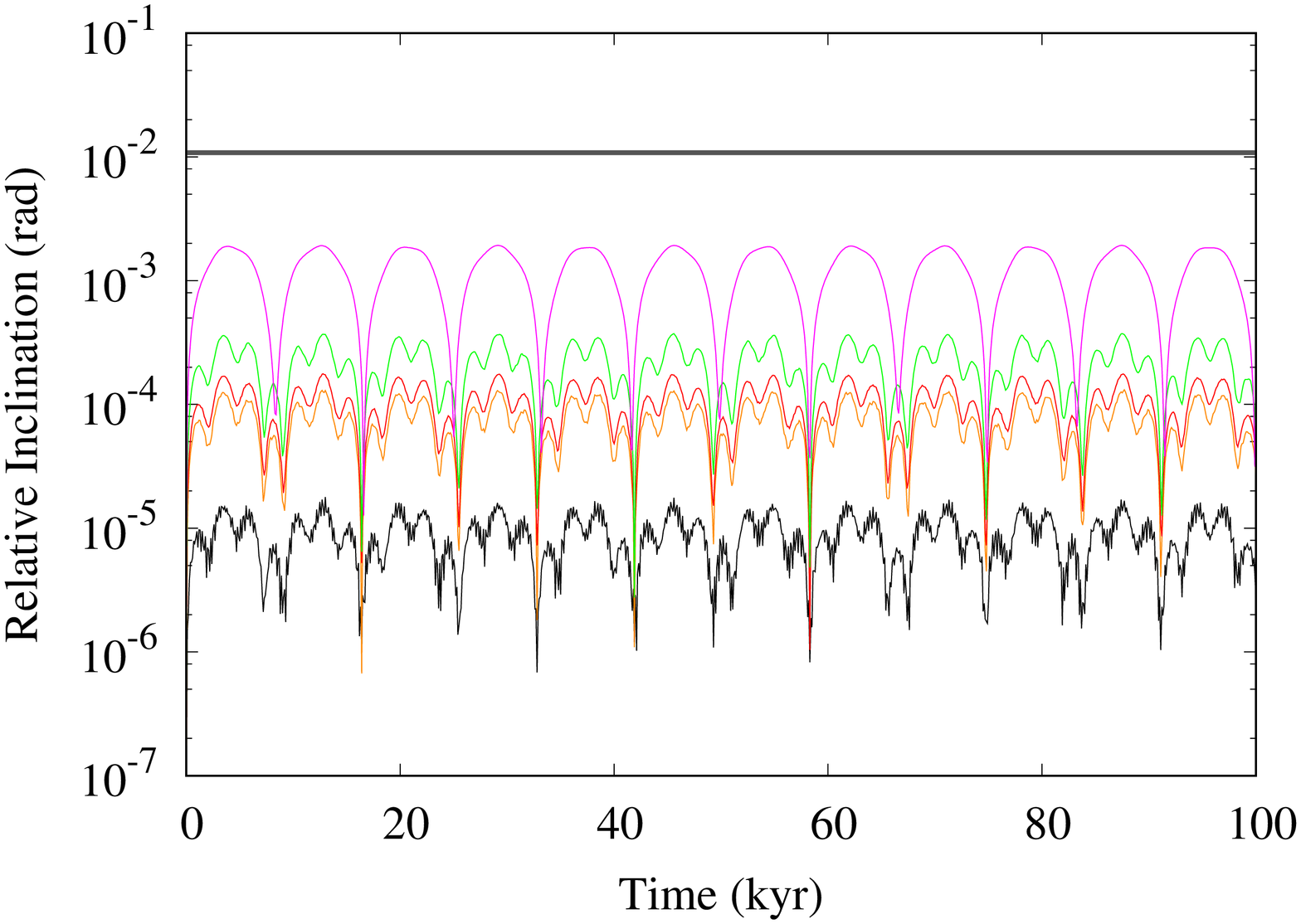}
\includegraphics [height = 3.3 in, angle = 270 ]{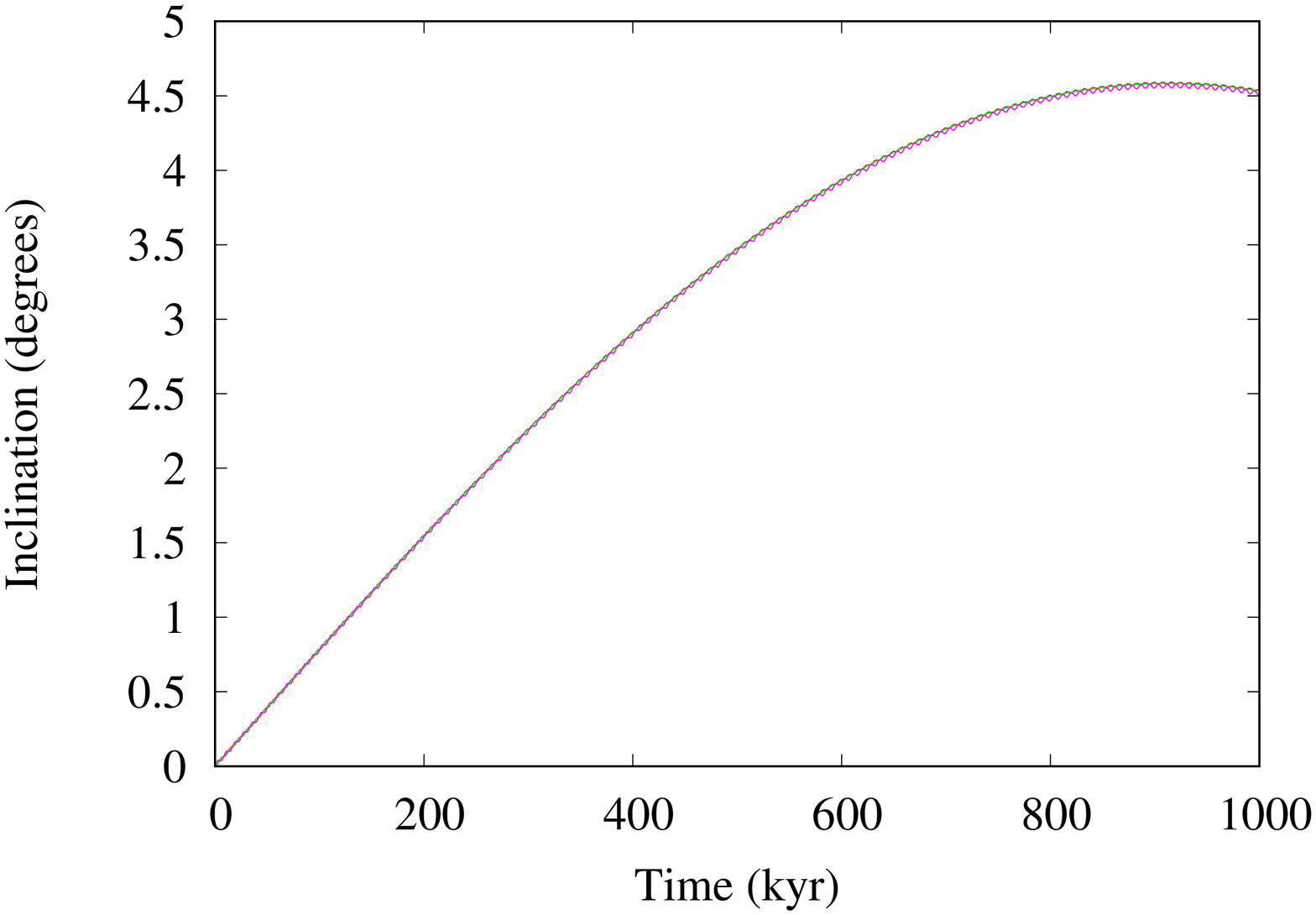}
\includegraphics [height = 3.3 in, angle = 270 ]{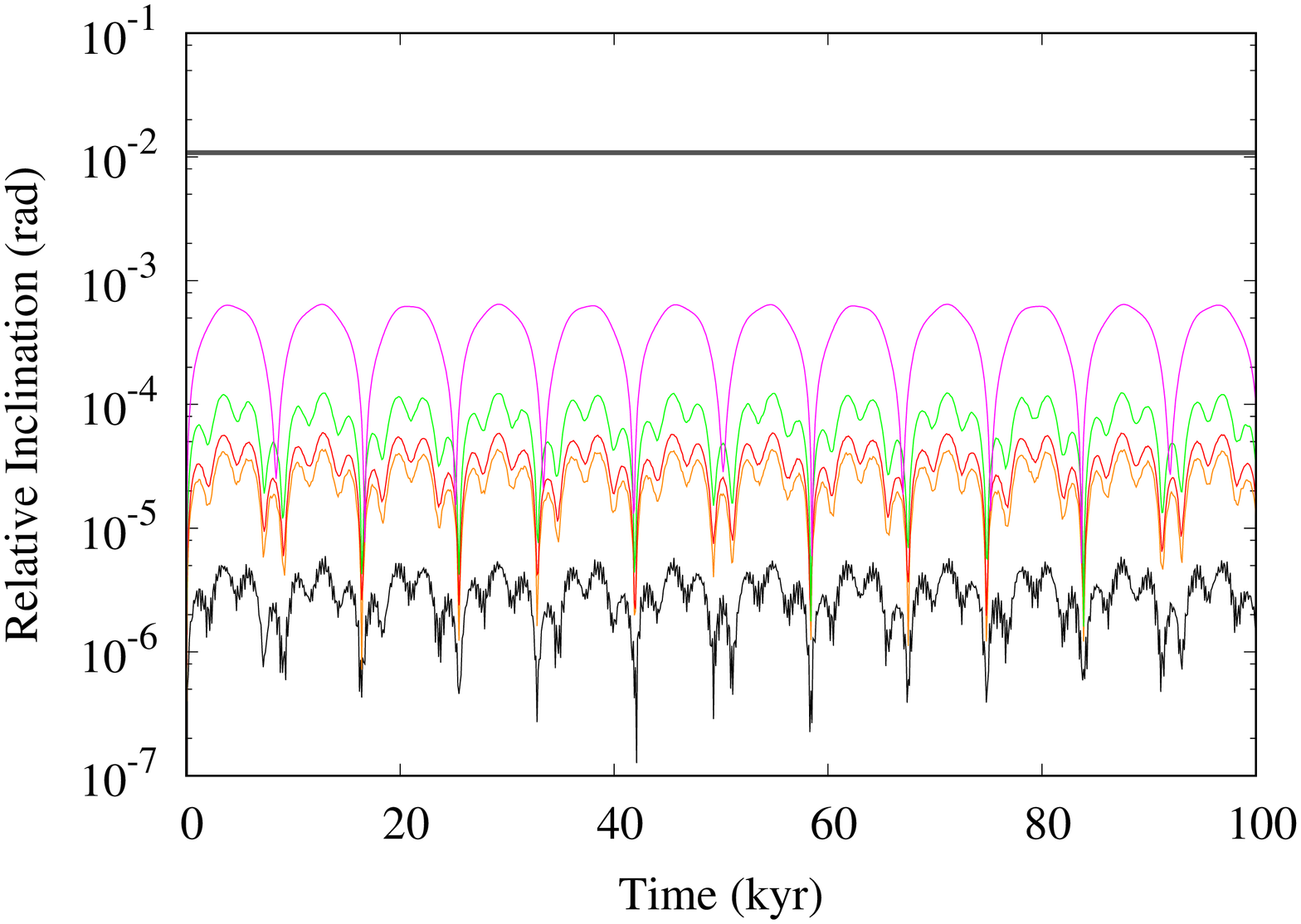}
\caption{Inclinations and relative inclinations for the six known planets orbiting Kepler-11, due to perturbations by 0.3 M$_{J}$ planet (top panels) or an 0.1 M$_{J}$ planet (bottom panels) at 3.0 AU with an initial 3 degree inclination. The left panels show the inclinations of all six transiting planets over time relative to the initial orbital plane of Kepler-11 b (in degrees). The right panels show inclinations relative to the contemporaneous orbital plane Kepler-11 b (in radians). The maximum relative inclination increases with increasing orbital distance from Kepler-11 b in color to Kepler-11 g in magenta. The grey horizontal line in the right panels mark $(R_{\star}/a)$ for Kepler-11 g. When all planets are below the grey line, they are considered co-planar or `likely co-transiting' for some observers. }
\label{fig:0.3Mj} 
\end{figure}

Finally, we checked the effect of a reduced mass of the distant perturber on the co-planarity of the transiting planets to determine at what distances a lower mass perturber could be ruled unlikely. In Figure~\ref{fig:0.3Mj} we plot the temporal variations of the mutual inclinations of the six known planets with a 0.3 M$_{Jup}$ and a 0.1 M$_{Jup}$ perturber at 3.0 AU with a 3$^{\circ}$ initial inclination. In this case, the timescale of the secular cycle in inclination increases from 250 kyr for a 1 M$_{Jup}$ perturber to $\sim$ 700 kyr for a 0.3 M$_{Jup}$ perturber and 2 Myr for a 0.1 M$_{Jup}$ perturber, although the timescale is irrelevant for our purposes. Most importantly, the mutual inclinations between the inner six never exceed the threshold to make their co-transiting configuration unlikely for the lower masses. 

We compared the maximum mutual inclinations caused by the perturber (at the same distance $a_p$) for the three choices of mass, shown in the right panels of Figure~\ref{fig:coplanar} and Figure~\ref{fig:0.3Mj}. The results are consistent with those of \citet{Lai2016}, as our maximum mutual inclination approximately scales with the perturbing mass $m_p$, although the high multiplicity of Kepler-11 makes the peak mutual inclination vary slightly between cycles.

\begin{figure}[h!]
\includegraphics [height = 4.5 in, angle = 270 ]{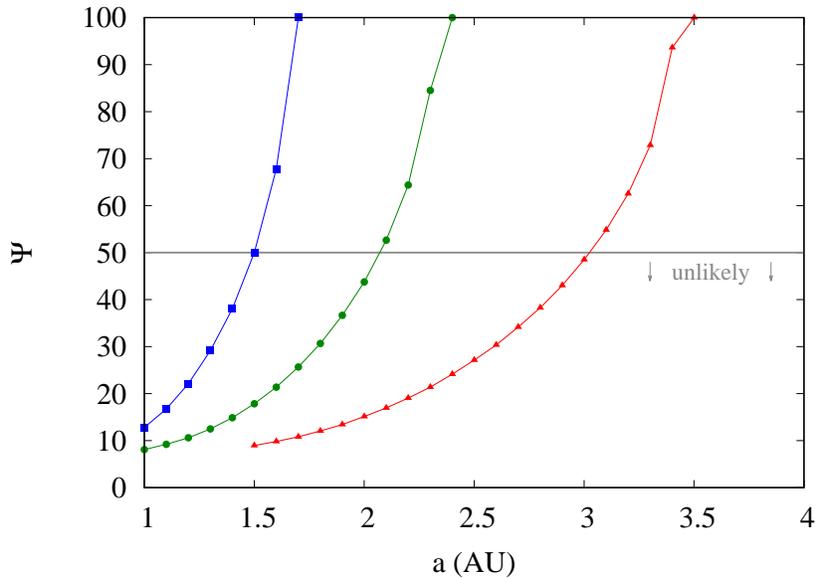}
\caption{The percentage of the time $\Psi$ that the transiting planets are ``coplanar" over an 8 Myr period, with a perturber inclined at 3$^{\circ}$ of mass 1 M$_{Jup}$ (red triangles), 0.3  M$_{Jup}$ (green circles), or 0.1  M$_{Jup}$ (blue squares). 
\label{fig:lowmasslimit} }
\end{figure}

Figure~\ref{fig:lowmasslimit} shows the distances at which planets on a 3$^{\circ}$ inclination with different masses 1 M$_{Jup}$, 0.3 M$_{Jup}$, or 0.1 M$_{Jup}$ can be ruled unlikely. We found that a Saturn-mass (0.3 M$_{Jup}$) is unlikely within 2 AU, similar to the RV constrain on a potential Jupiter-mass planet. A lower mass planet (0.1 M$_{Jup}$)
is unlikely within 1.5 AU. This is a tighter constraint than our TTV models provided, where the model fit of the TTVs was significantly degraded by a perturber of 0.1 M$_{Jup}$ only within 1 AU (see Figure~\ref{fig:TTVsJ}). 
 
\section{Discussion and Conclusions}
We define a system as unlikely if all six transiting planets of Kepler-11 would be co-transiting to a distant observer less than 50$\%$ of the time if Kepler-11 g is transiting. We have determined that the minimum distance out to which a Jovian mass planet at Kepler-11 is unlikely is 3.0 AU with a moderate inclination of 3$^{\circ}$. 

A Jovian mass planet at higher inclination can be ruled out to greater orbital distances.  In principle, this distance would be reduced if an undetected planet were orbiting between Kepler-11 f and Kepler-11 g. However, transit timing data makes the presence of a planet $\gtrsim$3 M$_\oplus$ in mass unlikely, and a planet less massive than 3 M$_\oplus$ has only a moderate effect on the coplanarity of the known planets. On the other hand, if a Jovian-mass planet is detected within an orbital distance of 3.0 AU from the star, then the most likely scenarios to explain the current co-transiting configuration of Kepler-11 b-g are either a low inclination of the Jovian mass planet compared to the transiting six, or a planet between Kepler-11 f and g which would enhance the co-planarity of the inner system. 

Due to the faintness of Kepler-11, very limited RV observations are available, although the data in hand are able to rule out a planet  with $m\sin i = M_{Jup}$ within 1.93 AU from the current data \citep{Weiss2016}. This constraint is significantly weaker than our nominal result. Therefore, these dynamical models provide the tightest constraints yet available on the presence of a putative Jovian-mass planet orbiting Kepler-11 beyond Kepler-11 g. 

We thank Soko Matsumura for comments which improved this paper. D.J. acknowledges the support of the University of the Pacific. We also acknowledge support from NASA Exoplanets Research Program awards \#NNX17AC23G and \#NNX15AE21G. This work was partially supported by funding from the Pennsylvania State University's Office of Science Engagement and Center for Exoplanets and Habitable Worlds.  The Center for Exoplanets and Habitable Worlds is supported by the Pennsylvania State University, the Eberly College of Science, and the Pennsylvania Space Grant Consortium. Portions of this research were conducted with Advanced CyberInfrastructure computational resources provided by The Institute for CyberScience at The Pennsylvania State University (http://ics.psu.edu). We thank the \textit{Kepler} mission for the extraordinary dataset which continues to advance our understanding of exoplanetary systems.  This study benefitted from collaborations and/or information exchanged within NASA's Nexus for Exoplanet System Science (NExSS) research coordination network sponsored by NASA's Science Mission Directorate. 


\end{document}